\documentclass[review]{elsarticle}

\usepackage{lineno,hyperref}
\modulolinenumbers[5]

\journal{International Journal of Heat and Mass Transfer}

\usepackage{amsmath,amssymb}
\usepackage{graphicx}
\usepackage{dcolumn}
\usepackage{bm}
\usepackage[utf8]{inputenc}
\usepackage[T1]{fontenc}
\usepackage{mathptmx}
\usepackage{makecell}

\begin{document}

\begin{frontmatter}

\title{Developed and Quasi-Developed Macro-Scale Heat Transfer in Micro- and Mini-Channels with Arrays of Offset Strip Fins Subject to a Uniform Heat Flux}

\author[1,2,3]{A. Vangeffelen\corref{mycorrespondingauthor}}
\cortext[mycorrespondingauthor]{Corresponding author}
\ead{arthur.vangeffelen@kuleuven.be.}
\author[1,2,3]{G. Buckinx}
\author[2,3]{C. De Servi}
\author[1,3]{M. R. Vetrano}
\author[1,3]{M. Baelmans}
\address[1]{Department of Mechanical Engineering, KU Leuven, Celestijnenlaan 300A, 3001 Leuven, Belgium}
\address[2]{VITO, Boeretang 200, 2400 Mol, Belgium}
\address[3]{EnergyVille, Thor Park, 3600 Genk, Belgium}




\begin{abstract}
In the present work, we examine to what degree the heat transfer can be described as developed on a macro-scale level in typical micro- and mini-channels with offset strip fin arrays subject to a uniform heat flux, considering flow entrance and side-wall effects. 
Full-scale numerical heat transfer simulations are conducted to determine the extent of the developed macro-scale heat transfer region within the arrays. 
We find that the onset point of developed heat transfer increases linearly with the Péclet number and channel width. 
However, the thermal development lengths remain limited relative to the overall channel length. 
Therefore, the local macro-scale heat transfer coefficient can be modeled by developed Nusselt number correlations with discrepancies below 25\% in both the developed and developing heat transfer regions. 
We observe that quasi-developed heat transfer prevails over nearly the entire entrance region of the channel and significantly contributes to the main heat transfer characteristics, particularly the eigenvalues and amplitudes of the dominant temperature modes.
Additionally, we analyze the impact of channel side walls on the temperature field's periodicity and the macro-scale temperature profile, which we characterize through an effective heat transfer coefficient. 
Our comprehensive numerical data covers various fin height-to-length ratios up to 1, fin pitch-to-length ratios up to 0.5, and channel aspect ratios ranging from 1/5 to 1/17, encompassing Reynolds numbers from 28 to 1224. 
Two sets of Prandtl number and thermal conductivity ratio are investigated, corresponding to the combinations of copper/air, and copper/water. 
\end{abstract}


\end{frontmatter}


\newpage
\clearpage

\section*{\label{sec:nomenclature}Nomenclature}
\vspace{-2mm}
\begin{table}[ht!]
    \begin{tabular}{ll}
        \textit{Latin symbols} & \\
        $c$ & specific heat capacity, J/(kg K) \\
        $\boldsymbol{B}_{fm}$ & tensor that maps $\mathrm{\nabla{T}}$ onto $\nabla \langle T \rangle_{m}^{f}$, - \\
        $\boldsymbol{B}_{sm}$ & tensor that maps $\mathrm{\nabla{T}}$ onto $\nabla \langle T \rangle_{m}^{s}$, - \\
        $\boldsymbol{D}$ & thermal dispersion source, m/(s K) \\
        $\boldsymbol{e}$ & unit vector, - \\
        $h$ & fin height, m \\
        $h_{\text{unit}}$ & developed heat transfer coefficient, W/(m\textsuperscript{3} K) \\
        $h_{b}$ & macro-scale heat transfer coefficient for bottom plate interface, W/(m\textsuperscript{3} K) \\
        $h_{fs}$ & macro-scale heat transfer coefficient for fluid-solid interface, W/(m\textsuperscript{3} K) \\
        $I$ & identity tensor, - \\
        $k$ & thermal conductivity, W/(m K) \\
        $\boldsymbol{K}_{d}$ & effective thermal dispersion tensor, m\textsuperscript{2}/s \\
$\boldsymbol{K}_{fs}$ & effective interfacial heat transfer tensor, W/(m\textsuperscript{2} K) \\
        $\boldsymbol{K}_{t}$ & effective thermal tortuosity tensor, - \\
        $l$ & fin length, m \\
        $\boldsymbol{l}$ & lattice vector of unit cell, m \\
        $l_{j}$ & lattice size of unit cell, m \\
        $L_{j}$ & size of channel domain along $\boldsymbol{e}_{j}$, m \\
        $m$ & normalized weighting function, - \\
        $\boldsymbol{n}$ & unit normal vector, - \\
        $N_{j}$ & number of unit cells along $\boldsymbol{e}_{j}$, - \\
        $Nu$ & Nusselt number, - \\
        $Pe$ & Péclet number, - \\
        $Pr$ & Prandtl number, - \\
        $q$ & heat flux, W/m\textsuperscript{2} \\
        $Re$ & Reynolds number, - \\
        $s$ & fin pitch, m \\
        $s_{0}$ & length of inlet region in front of fin array, m \\
        $s_{N}$ & length of outlet region after fin array, m \\
        \end{tabular}
\end{table}

\begin{table}[ht!]
    \begin{tabular}{ll}
        $t$ & fin thickness, m \\
        $T$ & temperature field, K \\
        $\mathrm{\nabla{T}}$ & developed temperature gradient, K/m \\
        $\boldsymbol{u}$ & velocity field, m/s \\
        $U_{\text{dev}}$ & developed macro-scale velocity, m/s \\
        $\boldsymbol{y}$ & coordinate relative to center of filter window, m \\
         & \\
    \end{tabular}
\end{table}

\begin{table}[ht!]
    \begin{tabular}{ll}
        \textit{Greek symbols} & \\
        $\gamma$ & phase indicator function, - \\
        $\Gamma$ & interface \\
        $\delta$ & Dirac indicator function, - \\
        $\epsilon$ & porosity, - \\
        $\epsilon_{0}$ & perturbation size, - \\
        $\Theta$ & temperature mode amplitude, K \\
        $\kappa$ & slope of macro-scale temperature difference profile, 1/m \\
        $\lambda$ & mode eigenvalue, - \\
        $\mu$ & dynamic viscosity, Pa s \\
        $\xi$ & non-dimensional macro-scale profile, - \\
        $\rho$ & density, kg/m\textsuperscript{3} \\
        $\boldsymbol{\psi}$ & vector field that maps $\mathrm{\nabla{T}}$ onto $T^{\star}$, m \\
        $\boldsymbol{\Psi}$ & vector field that maps $\mathrm{\nabla{T}}$ onto $\Theta$, m \\
$\Psi_{0}$ & closure variable amplitude, m \\
        $\Omega$ & domain \\
         & \\
    \end{tabular}
\end{table}

\begin{table}[ht!]
    \begin{tabular}{ll}
        \textit{Subscripts} & \\
        $b$ & related to the channel bottom plate / bulk average \\
        dev & developed contribution\\
        $e$ & extension \\
        $f$ & restricted to the fluid \\
        $fs$ & related to the fluid-solid interface \\
        in & related to the channel inlet \\
        $m$ & filtered by normalized weighting function $m$ \\
        out & related to the channel outlet \\
        periodic & related to the periodically developed regime \\
        predev & developing contribution \\
        quasi-dev & related to the quasi-developed regime \\
        quasi-periodic & related to the quasi-periodically developed regime \\
        ref & used as reference \\
        $s$ & restricted to the solid \\
        sides & related to the channel side walls \\
        $t$ & related to the channel top plate \\
        T & related to the temperature \\
        unit & unit cell \\
        uniform & related to the uniform macro-scale regime \\
         & \\
        \textit{Superscripts} & \\
        $\star$ & spatially periodic contribution \\
        $+$ & dimensionless \\
        $f$ & restricted to the fluid \\
        $s$ & restricted to the solid \\
    \end{tabular}
\end{table}

\newpage
\clearpage

\section{\label{sec:intro}Introduction}

As energy systems, industrial processes, electronics and data centers demand ever increasing heat power densities, researchers have focused on developing more compact heat transfer devices.
Over the past two decades, micro- and mini-channels with arrays of periodic fins have been shown to be very suitable for constructing highly-compact energy-efficient heat transfer devices \cite{kandlikar2005heat,khan2006role,izci2015effect,yang2017heatpin}. 
Particularly, micro- and mini-channels with periodic arrays of offset strip fins are frequently proposed as a solution, given that this fin type allows for high heat transfer coefficients and relatively low pressure drops. 
For that reason, they have been employed for the cooling of microelectronics \cite{bapat2006thermohydraulic, yang2007advanced, hong2009three}, the recovery of waste heat in compact gas turbines \cite{do2016experimental,nagasaki2003conceptual}, the refrigeration and liquefaction in cryogenic systems \cite{yang2017heat,jiang2019thermal}, and the heating of air flows in solar collectors \cite{yang2014design,pottler1999optimized}. 

Our previous studies \cite{vangeffelen2021friction,vangeffelen2022nusselt} highlighted that the flow regime in micro- and mini-channels with arrays of offset strip fins is mainly laminar and steady \cite{bapat2006thermohydraulic,yang2007advanced,hong2009three,do2016experimental,nagasaki2003conceptual,yang2017heat,jiang2019thermal,yang2014design,pottler1999optimized,tuckerman1981high}. 
The corresponding heat transfer regime in these channels is characterized by the flow's Prandtl number, which is usually 0.7 or 7 since air and water are the most common working fluids. 
Nevertheless, the Prandtl number of cryogenic fluids like helium, hydrogen, and nitrogen, both in the gaseous and liquid state, is of the same order of magnitude as that of air \cite{yang2017heat,jiang2019thermal}. 
Furthermore, the heat transfer regime in micro- and mini-channel applications is commonly investigated under the assumption of a steady and uniform heat flux at the channel wall. 
This boundary condition is considered to represent well the heat transfer process in microelectronics cooling systems, balanced counter-flow heat recuperators, as well as solar air heaters, and heat exchangers in cryogenic systems \cite{jiang2019thermal,yang2014design,pottler1999optimized,shah1978laminar,renfer2013microvortex,xia2017micro,zhang2011convective,gong2020heat,priyam2016thermal}. 

To model the flow and heat transfer in micro- and mini-channels with large arrays of periodic solid structures, an analysis based on a single unit cell of the array is often preferred \cite{liang2022fluid,odele2022performance}. 
After all, compared to a Direct Numerical Simulation (DNS) of the detailed heat transfer within the entire channel, a unit-cell approach significantly reduces the required computational resources \cite{kim2010thermoflow}.
In the scientific literature, two main theoretical frameworks can be distinguished that enable us to characterize the heat transfer regime through a unit-cell simulation. 

The first framework  relies on the assumption  that the heat transfer regime is periodically developed.
That way, the temperature distribution in the channel can be obtained on a single unit cell by solving the periodic flow and heat transfer equations formulated by Patankar et al. \cite{patankar1977fully}.
These equations not only govern the periodic part of the temperature field, which is similar within every unit cell, but also the overall temperature gradient over each unit cell.
This information allows us to obtain the total heat transferred between the fluid and the fins for any given flow rate.

The second framework treats the array of solid structures as a porous medium, allowing for the use of the volume-averaging technique (VAT) \cite{whitaker1996forchheimer,quintard1997two} to determine the volume-averaged or \textit{macro-scale} temperature field in the channel. 
In this approach, the  macro-scale heat transfer rate at the fluid-solid interface is modelled as proportional to the difference in macro-scale temperatures of the fluid and solid, through the introduction of an interfacial heat transfer coefficient.
This interfacial heat transfer coefficient is the solution to a closure problem that governs the local deviations from the macro-scale temperature field within the unit cell \cite{whitaker1996forchheimer,quintard1997two}. 
In practice, however, this closure problem is rarely solved, as it long has been assumed equivalent to the periodic heat transfer equations formulated by Patankar et al. \cite{patankar1977fully,nakayama2004numerical,degroot2011closure}. 

Yet, both frameworks can only be considered theoretically equivalent when specific weighting functions are selected to define the macro-scale temperature of the fluid and solid.
In most cases, the equivalence of both frameworks  requires the use of a double volume-averaging operation, as Buckinx and Baelmans (\cite{buckinx2015multi,buckinx2015macro,buckinx2016macro,buckinx2017macro}) demonstrated by 
extending the weighted volume-averaging technique proposed by Quintard and Whitaker \cite{quintard1994transport1,quintard1994transport2,quintard1994transport3,quintard1994transport4,quintard1997two,davit2017technical}.
The double volume-averaging operation not only leads to a physically meaningful macro-scale description of the developed flow and heat transfer regimes, but it also gives rise to mathematically exact and spatially constant heat transfer coefficients obtainable from a unit-cell simulation. 

In our previous works \cite{vangeffelen2021friction, vangeffelen2022nusselt, vangeffelen2023macro}, we have studied the periodically developed flow and heat transfer regime in micro- and mini-channels with offset strip fin arrays subject to a uniform heat flux. 
Specifically, we have determined the macro-scale pressure gradient and macro-scale heat transfer coefficient for the periodically developed regime, which we correlated in the form of a dimensionless friction factor and Nusselt number as a function of the Reynolds number, the geometrical parameters and the material properties. 
Moreover, using DNS, we have investigated to what extent the actual flow in the entire channel can be described as developed from a macro-scale perspective \cite{vangeffelen2023macro}.
Notably, we have found that the flow becomes quasi-periodically developed close to the channel inlet and quickly decays to a fully periodic regime for a wide range of Reynolds numbers and geometrical parameters.
As a result, the macro-scale flow is essentially developed over the entire channel, implying that our friction factor correlation \cite{vangeffelen2021friction} accurately predicts the overall pressure drop over the channel. 
These observations align with the work of Feppon \cite{feppon:hal-04443652}, which mathematically proves the existence of quasi-periodically developed solutions for Stokes flow in finite periodic channels. 
It also presents an alternative macro-scale flow description based on formal two-scale asymptotic expansions. 

Nevertheless, at present, it remains unclear how accurately our Nusselt number correlation \cite{vangeffelen2022nusselt} predicts the macro-scale heat transfer rate over the entire channel, or the overall temperature difference between the channel inlet and outlet. 
Ultimately, we still do not know at which distance from the channel inlet the heat transfer can be regarded as periodically developed in micro- and mini-channels with arrays of offset strip fins. 


In the literature, very few studies treat the development of the heat transfer regime in channel flows. 
Most of the available studies consider the development of the temperature field in two-dimensional channels without solid structures. 
As they start from the ansatz that the flow is fully developed, i.e., Poiseuille flow, their main target is to solve the so-called Graetz-Nusselt problem for the channel's specific cross-sectional geometry \cite{graetz1882ueber,graetz1885warmeleitungsfahigkeit,nusselt1910abhangigkeit}. 
Hereto, two solution methods are commonly applied, both yielding the thermal development length, as well as the incremental heat transfer rate in the channel entrance region. 

The first solution method, proposed by Graetz, consists of applying separation of variables to the developing temperature field, so that the separated problem becomes solvable via Sturm-Liouiville theory \cite{graetz1882ueber}. 
This means that the solution for the temperature field is obtained as an infinite series of eigenvalues and eigenfunctions for the Graetz-Nusselt problem. 

A second solution method is the Lévêque approach, which supplements the Graetz method with a similarity transformation. 
The purpose of the similarity transformation is to circumvent the need for an increasingly large number of terms in the series solution for the temperature field to obtain a good accuracy near the channel inlet \cite{leveque1928lois,awad2010heat}. 

Also, numerical methods have been employed to study the development of the heat transfer regime. 
Such methods usually rely on a finite-difference or finite-volume discretization to solve the thermal energy equation  \cite{grigull1965thermischer,montgomery1966laminar,tay1971application}. 
As a matter of fact, numerical methods have been the only means so far to determine the developing temperature field in a simultaneously developing flow.
Especially when the flow is turbulent instead of laminar, or the considered channel geometry is three-dimensional instead of two-dimensional, numerical methods have found widespread application, though only for channels without solid structures \cite{hwang1964finite,montgomery1968laminar,nguyen1993incremental,sparrow1986numerical}. 

On the contrary, no analytical or numerical studies exist in the literature on the development of heat transfer in channels with arrays of fins or other solid structures. 
It remains an open problem whether analytical methods for such channels can be formulated, as the available methods cannot easily be extended to deal with the more complex boundary conditions in three-dimensional arrays. 
The research gap in numerical studies is primarily due to the tremendously high computational cost of simulating the developing flow and temperature fields in large arrays \cite{kim2010thermoflow}. 
Even for steady laminar flows, which do not require turbulence modelling via Reynolds-Averaged Navier–Stokes (RANS) simulations or Large Eddy Simulations (LES) and allow the use of DNS, the main computational challenge remains the required mesh resolution, hence the global mesh size. 
In typical fin arrays, a minimal mesh size of the order of $10^5$ cells per unit cell is required for a sufficient spatial resolution of the flow and heat transfer \cite{kim2011correlations,vangeffelen2021friction,vangeffelen2022nusselt}. 
This makes DNS, as well as RANS or LES in case the developing flow would be turbulent, computationally prohibitive for such fin arrays. 
As a result, the general features of the developing flow and heat transfer in such channels are still barely understood, with the exception of the quasi-periodically developed flow regime \cite{buckinx2023arxiv}. 
The former regime is characterized by a single exponential velocity mode with a streamwise periodic amplitude and dominates over almost the entire entrance region in micro- and mini-channels with offset strip fins \cite{vangeffelen2023macro} and inline cylinders \cite{buckinx2022arxiv}. 
Although the existence of a similar quasi-periodically developed heat transfer regime has also been recently deduced \cite{buckinx2024arxiv}, its significance within the entire region of developing heat transfer has yet to be investigated, as empirical data is still lacking. 

Not only theoretical work, but also experimental work on the heat transfer development in channels with arrays of solid structures remains scarce. 
The most significant study, to the authors' knowledge, is by Dong et al. \cite{dong2007air}, who investigated the impact of thermal development on the thermo-hydraulic performance of offset strip fin arrays. 
They proposed a correlation for the Colburn j-factor, accounting for developing flow and heat transfer through a scaling factor based on the channel length-to-fin length ratio. 
However, this correlation primarily applies to air flows in the transitional regime and is specifically tailored to conventional offset strip fin channels with relatively high channel heights. 
Moreover, since the study involved high flow rates of hot water on the control side, it reflects a constant-wall-temperature condition. 
As a result, the findings of Dong et al. have limited applicability to our investigation of heat transfer in micro- and mini-channels with offset strip fins under a constant heat flux boundary condition. 

From the preceding literature survey we conclude that, currently, the importance of the heat transfer development in micro- and mini-channels with offset strip fin arrays is still unexplored. 
Therefore, the objective of this work is to theoretically and  numerically assess to what extent the heat transfer regime can be described as developed on a macro-scale level in common micro- and mini-channels with offset strip fins subject to a uniform heat flux. 
Specifically, we analyze the onset point of the developed heat transfer regime, as well as the onset point of quasi-developed heat transfer inside these channels. 
Hereto, we rely on DNS to resolve the entire temperature field in the simultaneously developing channel flow for Reynolds numbers ranging from 28 to 1224.
Both air and water are considered as working fluids, each in combination with a highly thermally conductive solid material, such as copper. 
Further, we assess the accuracy of the Nusselt number correlations from our preceding work \cite{vangeffelen2022nusselt} in the region where the heat transfer regime is still developing, since they are strictly speaking only valid for predicting the heat transfer coefficient in the developed region.
Concretely, we conduct a comparison with the actual macro-scale heat transfer coefficient along the channel, which we compute from the full temperature field through a discretized double volume-averaging operator. 
Finally, we investigate and quantify the influence of the channel's side walls on the developed macro-scale temperature field and the actual macro-scale heat transfer coefficient. 

The present work is structured as follows. 
Section \ref{sec:msheat_method} focuses on DNS, where we present the geometry of the channels and offset strip fins, the computational domain, the energy equations for steady conjugate heat transfer in the channel, as well as the numerical procedure to solve them. 
In Section \ref{sec:msheat_onsetMS}, the onset of the developed macro-scale heat transfer is determined, and the validity of our developed Nusselt number correlations is discussed. 
The onset point and features of the quasi-developed heat transfer regime are presented in Section \ref{sec:msheat_onsetquasi}. 
Lastly, Section \ref{sec:msheat_sidewall} treats the influence of the channel side wall on the macro-scale heat transfer.


\section{\label{sec:msheat_method} Channel geometry and (macro-scale) temperature equations}

\subsection{Geometry of the channel and fin array}

\begin{figure}[ht!]
\begin{center}
\includegraphics[scale = 0.75]{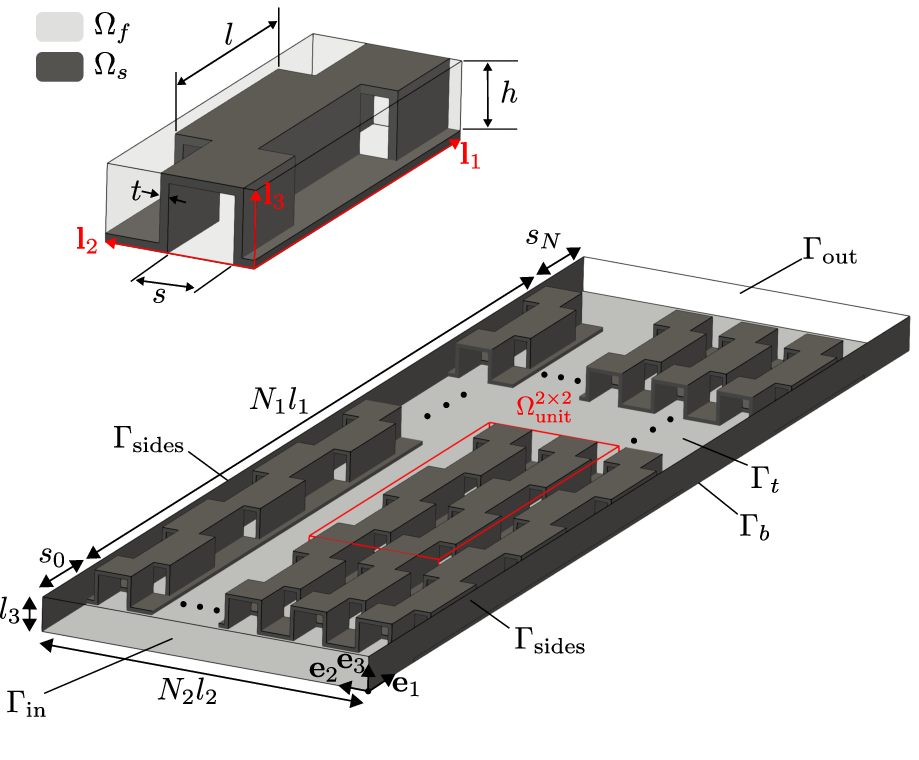}
\caption{\label{fig:msheat_geometrydns} Offset strip fin channel and unit cell domain}
\end{center}
\end{figure}

Figure \ref{fig:msheat_geometrydns} illustrates the channel domain $\Omega$ used for the direct numerical simulation (DNS) of the three-dimensional temperature field inside. 
The channel contains an offset strip fin array comprising $N_{1}$ unit cells along the main flow direction and $N_{2}$ unit cells along the lateral direction. 
The channel's total length $L_{1}$, total width $L_{2}$ and total height  $L_{3}$ are given by $L_{1} = (s_{0} + N_1 l_{1} + s_{N})$, $L_{2} = N_2 l_{2}$ and $L_{3} = l_{3}$, since $s_{0}$ and $s_{N}$ denote the distance between the fin array and channel inlet or outlet, respectively. 
We remark that the channel walls have zero thickness. 
Also, a single unit cell $\Omega_{\text{unit}}$ of the offset strip fin array is shown in Figure \ref{fig:msheat_geometrydns}. 
The geometry of the unit cell is determined by the fin length $l$, the fin height $h$, the lateral fin pitch $s$, and the fin thickness $t$. 
The porosity of the fin array is thus given by $\epsilon_{f} = hs/ \left[ \left( h+t \right) \left( s+t \right) \right]$. 
Each unit cell is spanned by three lattice vectors $\boldsymbol{l}_{1} = l_{1} \boldsymbol{e}_{1} = 2l \boldsymbol{e}_{1}$, $\boldsymbol{l}_{2} = l_{2} \boldsymbol{e}_{2} = 2(s+t) \boldsymbol{e}_{2}$ and $\boldsymbol{l}_{3} = l_{3} \boldsymbol{e}_{3} = (h+t) \boldsymbol{e}_{3}$ with respect to the normalized Cartesian vector basis $\{ \boldsymbol{e}_{j} \}_{j=1,2,3}$, as depicted in Figure \ref{fig:msheat_geometrydns}. 

In this work, the following array sizes have been selected for our DNS simulations: $N_1=20$, $N_2 \in \{5, 10, 13, 15, 17 \}$. 
Furthermore, an inlet and outlet region without fins over a length $s_0 = l_{1}$ and $s_N = 2.5 l_{1}$ have been added in front and behind the offset strip fin array. 
These parameter values are the same as those considered in our previous study on the macro-scale flow development in offset strip fin arrays \cite{vangeffelen2023developed}. 
They have been chosen such that our results are representative of typical geometries of micro- and mini-channels with offset strip fins \cite{vangeffelen2023developed}. 
We have verified that the number of fin rows $N_1=20$ is large enough to accurately assess the development lengths for both the flow and temperature fields. 
Besides, we have checked that the number of lateral unit cells $N_2$ results in true lateral periodicity of the flow field at a distance $l_2$ from the channel's sidewalls.



It is worth noting that if we use the fin length $l$ as the reference length, the non-dimensional geometrical parameters $h/l$, $s/l$, $t/l$, $s_0$, $s_N$, $N_{1}$, and $N_{2}$ completely specify of the channel geometry.

\vfill

\subsection{\label{sec:msheat_method_eqs} Temperature equations for steady conjugate heat transfer in a channel}
The temperature field $T$ in the channel $\Omega$, which equals $T_f$ in the fluid ($\Omega_{f}$) and $T_s$ in the solid material ($\Omega_{s}$), is obtained from the following energy conservation equations, as we neglect viscous dissipation and other heat sources: 
\begin{equation}
\begin{aligned}
  \rho_{f} c_{f} \nabla \cdot \left( \boldsymbol{u} T_{f} \right) &=  k_{f} \nabla^2 T_{f} & \text{in } \Omega_{f}, \\
  0 &= k_{s} \nabla^2 T_{s} & \text{in } \Omega_{s},
\end{aligned}
\label{eq:msheat_Temperature}
\end{equation}
Here, the thermal conductivity $k$, density $\rho$, and specific heat capacity $c$ of the fluid ($f$) and solid ($s$) are all assumed to be constant. 
The steady velocity field $\boldsymbol{u}$ has been obtained by solving the Navier-Stokes equations for an incompressible Newtonian fluid determined, as described in \cite{vangeffelen2023developed}. 

We consider the following boundary conditions for the temperature field in the channel. 
At the channel inlet $\Gamma_{\text{in}}$, a uniform inlet temperature $T_{\text{in}}$ is assumed. 
A uniform heat flux $q_{b}$ is imposed at the bottom wall of the channel, which is further subdivided into a part that bounds the fluid domain and a part that bounds the solid domain: $\Gamma_{b} = \Gamma_{bf} \cup \Gamma_{bs}$. 
At the channel outlet $\Gamma_{\text{out}}$, the top boundary $\Gamma_{t}$ and the side-wall boundaries $\Gamma_{\text{sides}}$, a no-heat-flux boundary condition, hence a zero Neumann boundary condition, is assumed. 
Finally, at the fluid-solid interface $\Gamma_{fs}$, we assume the continuity of the temperature field and the heat flux to adequately model the steady conjugate heat transfer between the fluid and solid. 
Mathematically, the former boundary conditions have the following form:
\begin{equation}
\begin{aligned}
  T_{f} \left( \boldsymbol{x} \right) &= T_{\text{in}} & &\text{in } \Gamma_{\text{in}}, \\ 
  - \boldsymbol{n} \cdot k_{f} \nabla T_{f} &= q_{b} & &\text{in } \Gamma_{bf}, \\
  - \boldsymbol{n} \cdot k_{s} \nabla T_{s} &= q_{b} & &\text{in } \Gamma_{bs}, \\
  - \boldsymbol{n} \cdot k_{f} \nabla T_{f} &= 0 & &\text{in } \Gamma_{\text{out}} \cup \Gamma_{t} \cup \Gamma_{\text{sides}}, \\
  - \boldsymbol{n} \cdot k_{s} \nabla T_{s} &= 0 & &\text{in } \Gamma_{t} \cup \Gamma_{\text{sides}}, \\
  T_{f} &= T_{s} & &\text{in } \Gamma_{fs}, \\
  - \boldsymbol{n}_{fs} \cdot k_{f} \nabla T_{f} &= - \boldsymbol{n}_{fs} \cdot k_{s}  \nabla T_{s} & &\text{in } \Gamma_{fs}. 
\end{aligned}
\label{eq:msheat_BoundaryConditionsHeat}
\end{equation}
We remark that in the boundary conditions (\ref{eq:msheat_BoundaryConditionsHeat}), the unit normal vector $\boldsymbol{n}$ at the exterior boundary $\Gamma = \partial \Omega$ points outward of the channel domain $\Omega$. 
As a result, the heat flux $q_{b}$ is negative when it is directed towards the channel $\Omega$. 
On the other hand, the unit normal vector $\boldsymbol{n}_{fs}$ at $\Gamma_{fs}$ points from the fluid domain $\Omega_{f}$ towards the solid domain $\Omega_{s}$. 

The former temperature equations (\ref{eq:msheat_Temperature}) and their boundary conditions (\ref{eq:msheat_BoundaryConditionsHeat}) are solved for two Prandtl numbers: $Pr_f=0.7$ and $Pr_f=7$, which correspond to air and water as the fluid.
Moreover, we have computed solutions for two thermal conductivity ratios: $k_{s}/k_{f}=10^4$ and $k_{s}/k_{f}=500$.
These ratios have been selected to represent the combinations of copper/air, and copper/water, respectively \cite{shah2003fundamentals}. 
The former thermal parameters are consistent with our previous study on the periodically developed heat transfer regime mini- and micro-channels with offset strip fins \cite{vangeffelen2022nusselt}.

\subsection{\label{sec:msheat_method_eqs_ms} Macro-scale temperature equations for steady conjugate heat transfer in a channel}
In agreement with the macro-scale descriptions from \cite{ buckinx2015multi, buckinx2015macro, buckinx2016macro, buckinx2017macro, quintard1994transport1, buckinx2022arxiv, vangeffelen2023developed}, we compute the macro-scale temperature fields by applying a double volume-averaging operator $\langle \; \rangle_{m}$ to the original temperature fields $T_f$ and $T_s$. 
This operator corresponds to the convolution product $\langle \phi \rangle_{m} = m \ast \phi$ in $\mathbb{R}^{3}$ with the  weighting function 
\begin{equation}
\label{eq:weightingfunction}
  m \left( \boldsymbol{y} \right) = \frac{1}{l_{3}} \text{rect} \left( \frac{y_{3}}{l_{3}} \right) \prod_{j=1}^{2} \frac{l_{j} - 2 | y_{j} | }{l_{j}} \text{rect} \left( \frac{y_{j}}{2 l_{j}} \right)\,.
\end{equation}
Because this weighting function is based on the normalized rectangle function \cite{weisstein2002rectangle}, its filter window is a double unit cell $\Omega_{\text{unit}}^{2 \times 2} \left( \boldsymbol{x} \right)$ with the same height as the channel \cite{buckinx2022arxiv}  (see Figure \ref{fig:msheat_geometrydns}). 
Consequently, the resulting macro-scale variables are two-dimensional fields: they vary only with the coordinate position in the midplane of the channel on which they are evaluated. 

In particular, we define the intrinsic macro-scale temperatures of the fluid and solid as $\langle T \rangle_{m}^{f} \triangleq \epsilon_{fm}^{-1} \langle T \gamma_{f} \rangle_{m}$ and $\langle T \rangle_{m}^{s} \triangleq \epsilon_{sm}^{-1} \langle T \gamma_{s} \rangle_{m}$, respectively. 
Here, we have introduced the fluid indicator  $\gamma_{f}$ and solid indicator $\gamma_{s}$, which are given by  $\gamma_{f}(\boldsymbol{x}) = 1 \leftrightarrow \boldsymbol{x} \in \Omega_{f}$, $\gamma_{f}(\boldsymbol{x}) = 0 \leftrightarrow \boldsymbol{x} \in  \mathbb{R}^{3} \setminus \Omega_{f} $ and $\gamma_{s}(\boldsymbol{x}) = 1 \leftrightarrow \boldsymbol{x} \in  \Omega_{s}$, $\gamma_{s}(\boldsymbol{x}) = 0 \leftrightarrow \boldsymbol{x} \in  \mathbb{R}^{3} \setminus \Omega_{s}$. 
Further, the weighted porosities are defined by $\epsilon_{fm} \triangleq \langle \gamma_{f} \rangle_{m}$ and $\epsilon_{sm} \triangleq \langle \gamma_{s} \rangle_{m}$. 

The intrinsic macro-scale temperature fields of the fluid and solid are governed by the following macro-scale energy conservation equations \cite{quintard1997two,buckinx2017macro}:
\begin{equation}
\begin{aligned}
  \rho_{f} c_{f} \nabla \cdot \left( \epsilon_{fm} \langle \boldsymbol{u} \rangle_{m}^{f} \langle T \rangle_{m}^{f} \right) &=  k_{f} \nabla^2 \left( \epsilon_{fm} \langle T \rangle_{m}^{f} \right) - \langle q_{b} \delta_{bf} \rangle_{m} - \rho_{f} c_{f} \nabla \cdot \boldsymbol{D} \\
  & \quad + k_{f} \nabla \cdot \langle \boldsymbol{n}_{fs} T_{f} \delta_{fs} \rangle_{m} + k_{f} \nabla \cdot \langle \boldsymbol{n} T_{f} \delta_{\text{sides,}f} \rangle_{m} \\
  & \quad + k_{f} \langle \boldsymbol{n}_{fs} \cdot \nabla T_{f} \delta_{fs} \rangle_{m} \\
  0 &= k_{s} \nabla^2 \left( \epsilon_{sm} \langle T \rangle_{m}^{s} \right) - \langle q_{b} \delta_{bs} \rangle_{m} \\
  & \quad - k_{s} \nabla \cdot \langle \boldsymbol{n}_{fs} T_{s} \delta_{fs} \rangle_{m} + k_{s} \nabla \cdot \langle \boldsymbol{n} T_{s} \delta_{\text{sides,}s} \rangle_{m} \\
  & \quad - k_{s} \langle \boldsymbol{n}_{fs} \cdot \nabla T_{s} \delta_{fs} \rangle_{m}\,.
\end{aligned}
\label{eq:msheat_TemperatureMS}
\end{equation}
In these energy equations, five closure terms can be distinguished. 
First, we have the macro-scale thermal dispersion source, $\boldsymbol{D} \triangleq \langle \boldsymbol{u} T \rangle_{m} - \epsilon_{fm} \langle \boldsymbol{u} \rangle_{m}^{f} \langle T \rangle_{m}^{f}$. 
This closure term represents the advective heat transfer that occurs by the local fluid velocity variations on top of the intrinsic macro-scale velocity $\langle \boldsymbol{u} \rangle_{m}^{f} \triangleq \epsilon_{fm}^{-1} \langle \boldsymbol{u} \gamma_{f} \rangle_{m}$. 
Secondly, we have the macro-scale thermal tortuosity, $ \langle \boldsymbol{n}_{fs} T_{f} \delta_{fs} \rangle_{m} = \langle \boldsymbol{n}_{fs} T_{s} \delta_{fs} \rangle_{m}$, which is based on the Dirac surface indicator $\delta_{fs}$ for $\Gamma_{fs}$. 
This closure term thus expresses a boundary integral of the temperature field over the curved (hence "tortuous") part of the fluid-solid interface contained within the filter window. 
Thirdly, we have the macro-scale interfacial heat transfer through conduction from the fluid to the solid along their interface $\Gamma_{fs}$ inside the filter window: 
\begin{equation}
     \langle q_{fs} \delta_{fs} \rangle_{m} = - k_{f} \langle \boldsymbol{n}_{fs} \cdot \nabla T_{f} \delta_{fs} \rangle_{m} = - k_{s} \langle \boldsymbol{n}_{fs} \cdot \nabla T_{s} \delta_{fs} \rangle_{m}. 
    \label{eq:msheat_closure_interf}
\end{equation}
Similarly, we recognize the macro-scale heat transfer imposed at the bottom of the channel, $\langle q_{b} \delta_{b} \rangle_m$, which can be split into the contributions  
$\langle q_{bf} \delta_{bf} \rangle_m$ and 
$\langle q_{bs} \delta_{bs} \rangle_m$, from the surfaces $\Gamma_{bf}$ and $\Gamma_{bs}$ respectively.
Finally, we have the two thermal tortuosity terms due to the channel's side walls, $k_{f} \nabla \cdot \langle \boldsymbol{n} T_{f} \delta_{\text{sides,}f} \rangle_{m}$ and $k_{s} \nabla \cdot \langle \boldsymbol{n} T_{s} \delta_{\text{sides,}s} \rangle_{m}$. 
They are based on the Dirac surface indicators  $\delta_{\text{sides,}f}$ and $\delta_{\text{sides,}s}$ corresponding to the part of $\Gamma_{\text{sides}}$ which bounds $\Omega_{f}$ and $\Omega_{s}$, respectively. 

It should be noted that the former macro-scale temperature equations are technically only valid when the overall temperature field $T$ is interpreted as an extended distribution of the form $T = T_{f}$ in $\Omega_{f}$, $T = T_{s}$ in $\Omega_{s}$ and $T = T_{e}$ in $\mathbb{R}^{3} \setminus \Omega$. 
In this work, the extension $T_{e}$ has been determined by extrapolation from the channel inlet, channel outlet, top boundary, and bottom boundary: $\forall s>0: T_{e} (\boldsymbol{x} +s\boldsymbol{n}) = T_{f} (\boldsymbol{x})$ when $\boldsymbol{x} \in ( \Gamma_{\text{in}} \cup \Gamma_{\text{out}} \cup \Gamma_{\text{t}} \cup \Gamma_{\text{b}} )$. 
Away from the channel side walls, we select $\forall s>0: T_{e} (\boldsymbol{x} +s\boldsymbol{n}) = 0$ when $\boldsymbol{x} \in \Gamma_{\text{sides}}$. 
In this manner, we avoid the necessity to include additional closure terms in (\ref{eq:msheat_TemperatureMS}), the so-called commutation errors, which arise due to discontinuous temperature jumps at the domain boundaries \cite{buckinx2017macro}. 

In line with the literature \cite{buckinx2015macro,buckinx2016macro}, we represent the macro-scale heat transfer rate at the fluid-solid interface by means of an interfacial heat transfer coefficient:
\begin{equation}
    h_{fs} \triangleq \epsilon_{fm}^{-1} \frac{\langle q_{fs} \delta_{fs} \rangle_m}{\langle T \rangle_{m}^{f} - \langle T \rangle_{m}^{s}} \,.
\label{eq:msheat_hfs}
\end{equation}
The latter is related to the non-dimensional Nusselt number  $Nu_{fs} \triangleq h_{fs} l^{2} / k_{f}$. 
In this work, we also make use of the following macro-scale heat transfer coefficient to represent the macro-scale heat transfer rate at the channel's bottom wall:
\begin{equation}
    h_{b}  \triangleq  \epsilon_{fm}^{-1} \frac{\langle q_{b} \delta_{b} \rangle_m}{\langle T \rangle_{m}^{f} - \langle T \rangle_{m}^{s}},
\label{eq:msheat_Nub}
\end{equation}
so we define $Nu_{b} \triangleq h_{b} l^{2} / k_{f}$. 




\subsection{Numerical procedure}

To solve the temperature equations (\ref{eq:msheat_Temperature})-(\ref{eq:msheat_BoundaryConditionsHeat}), we used the software package FEniCSLab, which was developed by G. Buckinx within the finite-element computing platform FEniCS \cite{AlnaesBlechta2015a}. 
A structured mesh was used for the spatial discretization of the offset strip fin channel, identical to the one employed for the flow simulations in our previous work \cite{vangeffelen2023developed}. 
The reason is that the developing velocity fields were also obtained from that work. 
The temperature field has been discretized by continuous Galerkin tetrahedral elements of the second order. 
Each temperature simulation required a computational time of at most 30 minutes on 10 nodes with 36 processors each (Xeon Gold 6140 2.3GHz with 192GB of RAM). 
The algebraic linear system of discretized temperature equations was solved iteratively using the generalized minimal residual method (GMRES) method with a Jacobi preconditioner \cite{petsc-web-page}. 
For all the cases considered in this work, a mesh-independence study has been performed, which indicated that the discretization error on the temperature profiles remains below 3\%. 
Finally, the same explicit finite-element integral operator as in \cite{vangeffelen2023developed} has been used for the double volume-averaging operations.
The former was implemented in FEniCSLab by G. Buckinx \cite{buckinx2022arxiv}.
For each temperature field, the calculation of the discrete double volume-averaging operation took about 6 hours on 10 nodes with 36 processors (Xeon Gold 6140 2.3GHz with 192GB of RAM).

\begin{figure}[ht]
\includegraphics[scale = 0.75]{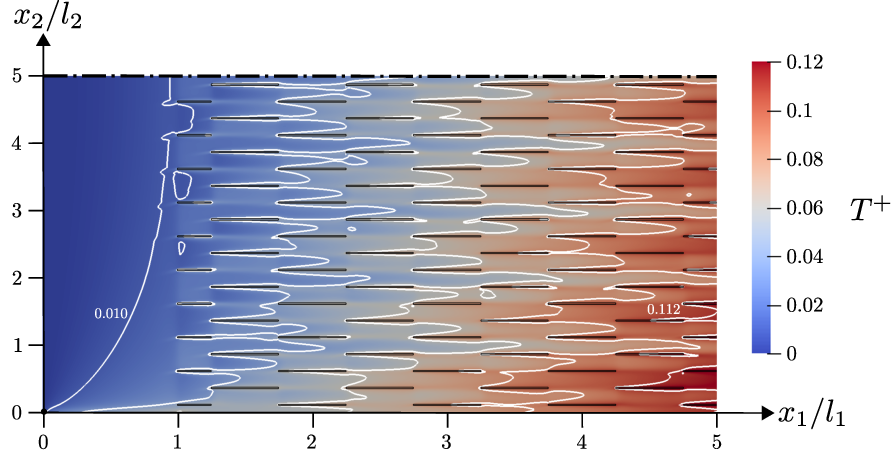}
\caption{\label{fig:isothermsPR7RE100T2H12S48N10}The non-dimensional temperature field $T^{+} \triangleq (T-T_{\text{in}}) k_f/(q_b l)$ in an offset strip fin channel for $Pr_f=7$, $Re_{b}=28$, $h/l=0.12$, $s/l=0.48$, $t/l=0.02$, $s_0=l_1$, $s_N=2.5 l_1$, $N_{1}=20$, and $N_{2}=10$. Its iso-lines are shown in half of the mid-plane spanned by $\boldsymbol{e}_{1}$ and $\boldsymbol{e}_{2}$. The black rectangles indicate the outlines of the offset strip fins.}
\end{figure}

Figure \ref{fig:isothermsPR7RE100T2H12S48N10} illustrates the simulated temperature field in an offset strip fin channel for a   Prandtl number $Pr_f=7$ and a Reynolds number $Re_{b} \triangleq \rho_{f} u_{b} (2 L_{3} ) / \mu_{f} = \rho_{f} u_{b} 2 (h + t) / \mu_{f} = 28$, based on the the bulk average velocity $u_{b}$ as defined in \cite{vangeffelen2023developed}. 
The considered geometry is specified by $h/l=0.12$, $s/l=0.48$, $t/l=0.02$, $s_0=l_1$, $s_N=2.5 l_1$, $N_{1}=20$, and $N_{2}=10$. 
The temperature profiles are visualized through iso-lines of the non-dimensional temperature $T^{+} \triangleq (T-T_{\text{in}}) k_f/(q_b l)$ in the mid-plane of the channel, spanned by $\boldsymbol{e}_{1}$ and $\boldsymbol{e}_{2}$. 
To ensure the clarity of the figure, only half of the region near the channel inlet is shown. 
It is clear that, after a short distance from the start of the fin array (at $x_1 \simeq 2 l_1$), the temperature profiles around the offset strip fins in the array appear to become qualitatively similar, despite the presence of some irregularities, which are artefacts resulting from the employed visualization software \cite{ParaView}. 
This similarity in the temperature field indicates the occurrence of the periodically developed heat transfer regime. \\

\newpage
\clearpage

\section{\label{sec:msheat_onsetMS}Onset of developed macro-scale heat transfer}

We first examine the onset point of the periodically developed heat transfer regime by analyzing the computed temperature fields of the air and water flows for various Reynolds numbers and channel geometries. 
This onset point also marks the onset of the developed macro-scale heat transfer regime, as the latter occurs at most a distance $l_1$ further downstream along the main flow direction $\boldsymbol{e}_{1}$. 
The reason for this is that $l_1$ is the filter radius of the double volume-averaging operator (\ref{eq:weightingfunction}), hence the length scale over which local temperature gradients are smoothed in our macro-scale description.

\subsection{\label{sec:msheat_onsetperiodic}Onset of periodically developed heat transfer}


As we consider a constant heat flux at the channel wall, the periodically developed heat transfer regime is characterized by a linearly varying temperature field, on which a streamwise periodic temperature part is superposed \cite{patankar1977fully,penha2012fully}:
\begin{equation}
\label{eq:msheat_Tdev}
T \simeq T_{\text{dev}} \triangleq \mathrm{\nabla{T}} \cdot \boldsymbol{x} + T^{\star}.
\end{equation}
Here, $\mathrm{\nabla{T}}$ is the spatially constant overall temperature gradient, while the periodic temperature part satisfies $T^{\star} \left( \boldsymbol{x} + \boldsymbol{l}_{1} \right) = T^{\star} \left( \boldsymbol{x} \right)$, just like the streamwise periodic velocity field $\boldsymbol{u} = \boldsymbol{u}^{\star}$ in this region.

Accordingly, we define the onset point $x_{\text{periodic,T}}$ of the periodically developed heat transfer regime as the streamwise coordinate $x_{1}$ after which the temperature field agrees with the periodically developed solution (\ref{eq:msheat_Tdev}) within 10\% of the overall temperature change over a single unit cell, so $ | T - T_{\text{dev}} | / | \mathrm{\nabla{T}} \cdot \boldsymbol{l}_{1} | \leqslant 0.1$ for $x_{1} \geqslant x_{\text{periodic,T}}$. 
The practical advantage of this definition lies in the fact the value of $x_{\text{periodic,T}}$ remains independent of the total array length. 
Moreover, due to the employed criterion for defining $x_{\text{periodic,T}}$, we can generally expect expression (\ref{eq:msheat_Tdev}) to hold within at least 1\% of the total temperature difference over the entire developed region. 
The explanation is that the developed region reasonably covers more than 10 unit cells in the streamwise direction, given the high number of fin rows in contemporary micro- and mini-channels \cite{vangeffelen2023developed}. 
Another interesting consequence of the chosen criterion is illustrated later in this section: 
For all the cases examined in this study, the local macro-scale heat transfer coefficient and local Nusselt number along the centerline of the channel are observed to deviate at most 5\% from their constant values in the developed region, when this region is identified as $x_{1} \geqslant x_{\text{periodic,T}}$. \\

\begin{figure}[ht!]
\centering
\begin{minipage}{.475\textwidth}
\hspace{-5mm}
\includegraphics[scale = 0.75]{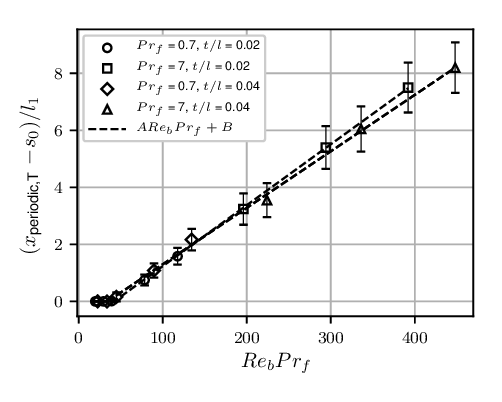}
\caption{\label{fig:msheat_xdevT_PE_t} Influence of the Péclet number on the onset of streamwise periodically developed heat transfer, when $N_{2}=10$, $h/l=0.12$, $s/l=0.48$ \newline}
\end{minipage}
\hfill
\begin{minipage}{.475\textwidth}
\hspace{-5mm}
\includegraphics[scale = 0.75]{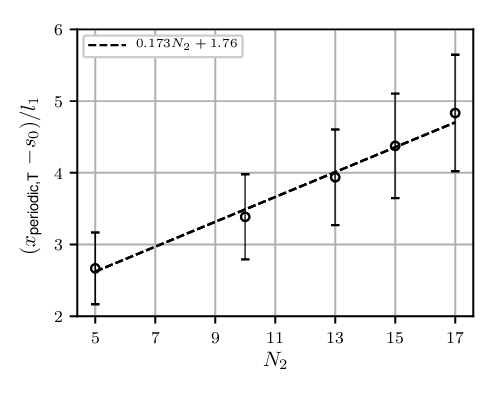}
\caption{\label{fig:msheat_xdevT_N2} Influence of the channel aspect ratio on the onset of streamwise periodically developed heat transfer, when $Pr_{f}=7$, $Re_{b}=28$, $h/l=0.12$, $s/l=0.48$, $t/l=0.02$}
\end{minipage}
\end{figure}

In Figure \ref{fig:msheat_xdevT_PE_t}, the onset point $x_{\text{periodic,T}}$ has been quantified for an array with 10 offset strip fins along the lateral direction, having a small height ($h/l=0.12$) and moderate spacing ($s/l=0.48$). 
In particular, the dependence of the onset point $x_{\text{periodic,T}}$ on the bulk Reynolds number $Re_{b}$ and the Prandtl number $Pr_{f}$ is illustrated. 
The error bars in the figure indicate the numerical uncertainty associated with discretization errors. 
Clearly, a linear relationship can be observed between the onset point $x_{\text{periodic,T}}$ and the Reynolds number $Re_{b}$. 
This linear relationship is inherited from the onset point of the periodically developed flow regime \cite{vangeffelen2023developed}.  
Indeed, before the temperature field can become developed, first the flow field itself must become (close to) developed, so both onset points will tend to move more downstream at a similar rate as $Re_{b}$ increases.

Figure \ref{fig:msheat_xdevT_PE_t} shows that the relation between the onset point $x_{\text{periodic,T}}$ and the Prandtl number $Pr_{f}$ can be reduced to a linear dependence on the Péclet number $Pe \triangleq Re_{b} Pr_{f}$. 
This appears to be true for all geometries and flow conditions studied in this work. 
We found, for instance, that the data in Figure \ref{fig:msheat_xdevT_PE_t} is captured by the following linear correlations within a maximum error of 5\%: $(x_{\text{periodic,T}} - s_{0}) / l_{1} \simeq 0.0214 Pe - 0.899$ when $t/l=0.02$, and $(x_{\text{periodic,T}} - s_{0}) / l_{1} \simeq 0.0198 Pe - 0.680$ when $t/l=0.04$. 
It should be noted though that the former linear correlations are only valid for $Pe > 42$ and $Pe > 34$, respectively.  
For smaller values of the Péclet number, the onset point approximately coincides with the start of the fin array: $x_{\text{periodic,T}} \simeq s_{0}$, as it can be seen in Figure \ref{fig:msheat_xdevT_PE_t}. 

The linear relationship between the onset point and Péclet number $x_{\text{periodic,T}} / l_{1} \simeq A Pe + B$, with $A$ and $B$ some constants, reflects that the distance over which the temperature field develops, is essentially determined by the rate at which thermal energy diffuses perpendicularly to the main flow direction, once the flow field has become developed. 
After all, the rate of thermal diffusion is given by $1/Pe$ when compared to the rate of thermal advection in the main flow direction.

The fact that $x_{\text{periodic,T}}$ increases with $Pe$ also indicates that the Prandtl number $Pr_{f}$ is a measure for the spatial distance over which the onset of developed heat transfer lags with respect to the onset of developed flow. 
This should be no surprise, since development essentially occurs through the diffusion of momentum and thermal energy in the directions perpendicular to the bulk flow, while the Prandtl number can be interpreted as the relative strength of thermal diffusion compared to momentum diffusion. 
For that reason, the onset point $x_{\text{periodic,T}}$ is located upstream of the onset point $x_{\text{periodic}}$ of the periodically developed flow regime when $Pr_{f} = 0.7$ \cite{vangeffelen2023developed}. 
On the contrary, when $Pr_{f} = 7$, $x_{\text{periodic,T}}$  and $x_{\text{periodic}}$ have a comparable magnitude, although their ratio is heavily dependent on the precise criterion used for defining $x_{\text{periodic,T}}$ and $x_{\text{periodic}}$. 

The observation that the onset point of the periodically developed heat transfer regime depends only on the Péclet number implies that this onset point is not influenced by the fluid's viscosity. 
This is a direct similarity between micro- or mini-channel with offset strip fins and other channels without solid structures. 
In fact, a linear scaling of the onset point with the Péclet number has been shown to occur in channels with a fixed hydraulic cross-section, not only when the heat transfer and flow regime are developing simultaneously \cite{everts2020laminar}, but also when the flow regime is already fully developed \cite{muzychka2002laminar,lee2006thermally,ma2021numerical}. 

Despite this similarity, it should be noted that the thermal development lengths $(x_{\text{periodic,T}} - s_{0})$ from this work cannot be obtained by simply re-scaling the flow development lengths from our previous work \cite{vangeffelen2023developed} with the Prandtl number: $x_{\text{periodic,T}} \neq x_{\text{periodic}} Pr_{f}$. 
This is a significant difference from what has been observed for flows in channels without solid structures \cite{everts2020laminar}, as well as external boundary-layer flows \cite{schlichting1961boundary}. 
Yet, we hypothesize that this difference can be attributed to the fact that in such types of flow, the streamwise diffusion of momentum and thermal energy is less important than in channels with fin arrays. 
In boundary-layer flows and channel flows of which the cross-sectional area does not change, streamwise diffusion of momentum and thermal energy is almost entirely overshadowed by streamwise advection of momentum and thermal energy. 
Consequently, the former's effect on the development lengths can be ignored because the characteristic length scale for streamwise advection is also quite large, typically comparable to the channel length. 
On the contrary, in channels containing a fin array, the characteristic length scale for streamwise advection is much smaller, often comparable to the spatial period $l_1$ of the array. 
The reason is that the steady flow patterns in a fin array, like wakes, recirculation zones, and vortices induced by the periodic variations in cross-sectional flow area are confined by the space between the periodic solid structures.

In Figure \ref{fig:msheat_xdevT_PE_t}, we have also illustrated the influence of the fin thickness-to-length ratio $t/l$ on the onset point of the periodically developed heat transfer regime. 
It can be seen that the thickness ratio $t/l$ has only a minimal influence on the onset point $x_{\text{periodic,T}}$. 
This is because the ratio $t/l$ has no significant impact on the onset point of periodically developed flow either \cite{vangeffelen2023developed}. 

Conversely, the aspect ratio of the channel $L_{3}/L_{2}=l_{3}/(N_{2} l_{2})$ does have a significant impact on the onset point, as Figure \ref{fig:msheat_xdevT_N2} shows. 
This figure shows that the onset point will move several unit cells more downstream when the aspect ratio of the channel decreases, hence when the number of unit cells in the lateral direction $N_{2}$ increases. 
The relationship between $x_{\text{periodic,T}}$ and $N_{2}$ is almost perfectly linear, as the straight lines in Figure \ref{fig:msheat_xdevT_N2} capture the data points with a maximum relative error of 5\%. 
This finding aligns with the observed linear relation between the flow development length and channel width in the same channels \cite{vangeffelen2023developed}. 
From a physical point of view, the linear scaling can be explained by noting again that thermal development occurs mostly due to lateral diffusion of thermal energy until all upstream perturbations from the developed temperature profile are leveled off. 
As the width of the channel corresponds to the distance over which these perturbations need to be transported via lateral diffusion, a higher channel width will result in a proportionally longer diffusion time and, consequently, a proportionally longer development length. 

Also, in rectangular channels without solid structures, the thermal development length increases for decreasing aspect ratios, although not necessarily in an inversely linear fashion \cite{muzychka2002laminar,lee2006thermally,ma2021numerical}. 
However, in those studies, the aspect ratio values are much larger than those in this work.
Therefore, a direct comparison cannot be made. 

\begin{figure}[h!]
\centering
\begin{minipage}{.475\textwidth}
\hspace{-5mm}
\includegraphics[scale = 0.75]{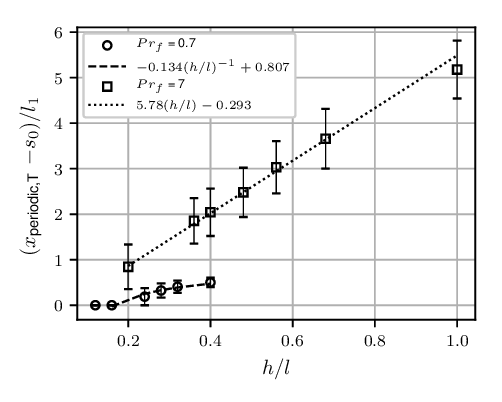}
\caption{\label{fig:msheat_xdevT_h} Influence of the fin height-to-length ratio on the onset of streamwise periodically developed heat transfer, when $Re_{b} l / (2 L_{3}) = \rho_{f} u_{b} l / \mu_{f} = 600$, $N_{2}=10$, $s/l=0.12$, $t/l=0.02$}
\end{minipage}
\hfill
\begin{minipage}{.475\textwidth}
\hspace{-5mm}
\includegraphics[scale = 0.75]{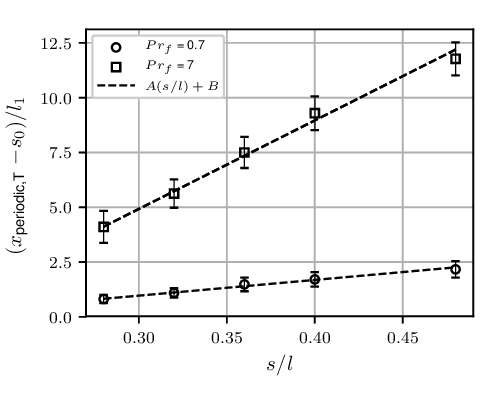}
\caption{\label{fig:msheat_xdevT_s} Influence of the fin pitch-to-length ratio on the onset of streamwise periodically developed heat transfer, when $Re_{b}=192$, $N_{2}=10$, $h/l=0.12$, $t/l=0.04$}
\end{minipage}
\end{figure}

Next to the aspect ratio, also the fin height-to-length ratio $h/l$ has a significant influence on the onset point $x_{\text{periodic,T}}$, as illustrated in Figure \ref{fig:msheat_xdevT_h}. 
In this figure, we have kept the bulk velocity $u_{b}$ and fin length $l$ constant to highlight the isolated effect of changing the fin height-to-length ratio $h/l$. 
Hereto, we have  specifically selected $Re_{b} l / (2 L_{3}) = \rho_{f} u_{b} l / \mu_{f} = 600$, which implies that the bulk Reynolds number $Re_{b}$ varies within the range of $(168, 1224)$ for $h/l \in (0.12, 1)$. 

We can distinguish two trends in Figure \ref{fig:msheat_xdevT_h}. 
First, we observe that the onset point $x_{\text{periodic,T}}$ moves downstream as the fin height-to-length ratio $h/l$ increases. 
This comes from the fact that the fin height $h$, just like the channel width $L_2$, is a distance over which upstream perturbations from the developed temperature profile need to be leveled off via thermal diffusion before developed heat transfer can occur. 
Therefore, we also observe a nearly linear relation between the onset point $x_{\text{periodic,T}}$ and fin height $h$ for intermediate values of $h/l$. 
Secondly, we observe that the onset point $x_{\text{periodic,T}}$ eventually becomes independent of the relative fin height $h/l$ upon further increasing $h/l$. 
This is a result of the temperature field becoming more two-dimensional and thus independent of $h/l$ for larger channel heights, following the flow field \cite{vangeffelen2022nusselt}. 
This second trend is, however, only visible for the lowest Prandtl number $Pr_{f} = 0.7$ in Figure \ref{fig:msheat_xdevT_h}. 
Indeed, for $Pr_{f} = 7$, the trend is expected to occur at much higher $h/l$ values, outside the range considered in Figure \ref{fig:msheat_xdevT_h}. 
We argue that when the Prandtl number $Pr_{f}$ increases, thermal diffusion becomes less important with respect to momentum diffusion so that the temperature field will become two-dimensional at larger values of $h/l$ \cite{vangeffelen2022nusselt}. 
Anyway, both trends are further supported by the empirical correlations displayed in Figure \ref{fig:msheat_xdevT_h}, which are able to predict the data for $h/l \geqslant 0.16$ and $h/l \geqslant 0.2$ when $Pr_{f} = 0.7$ and $Pr_{f} = 7$, respectively, within a relative error of 6\%.



Finally, Figure \ref{fig:msheat_xdevT_s} displays the influence of the fin pitch-to-length ratio $s/l$ on the onset point $x_{\text{periodic,T}}$. 
The influence of $s/l$ is again well predicted by a linear relation, since the following fitted curves are accurate to within 6\%:
$(x_{\text{periodic,T}} - s_{0}) / l_{1} \simeq 7.18 (s/l) - 1.19$ for $Pr_{f} = 0.7$ and $(x_{\text{periodic,T}} - s_{0}) / l_{1} \simeq 40.4 (s/l) - 7.19$ for $Pr_{f} = 7$.
The linearity of this relation resembles the linear trend between the onset point $x_{\text{periodic,T}}$ and relative width $N_{2}$, which suggests that the primary influence of $s/l$ consists of changing the transversal length over which diffusive transport must occur in the course of thermal development. 

In general, from the data shown in Figures \ref{fig:msheat_xdevT_PE_t}-\ref{fig:msheat_xdevT_s}, we conclude that for air, the thermal development length $(x_{\text{periodic,T}} - s_{0})$ typically remains below the streamwise length of two unit cells in offset strip fin micro- and mini-channels. 
Furthermore, the thermal development length for water remains smaller than 12 unit cell lengths for all cases considered in this work.
So, the short thermal development lengths are in agreement with the short flow development lengths observed in these channels \cite{vangeffelen2023developed}. 

We remark that, in addition to the onset point, we have also assessed the end point of the periodically developed heat transfer region in the former offset strip fin arrays. 
This end point $x_{\text{end,T}}$ has been computed as the streamwise coordinate $\boldsymbol{x}_{1}$ after which the temperature field deviates again from the periodically developed solution with more than 10\% of the temperature change over the unit cell: $ | T - T_{\text{dev}} | / | \mathrm{\nabla{T}} \cdot \boldsymbol{l}_{1} | \geqslant 0.1$ for $x_{1} \geqslant x_{\text{end,T}}$. 
For all the computed cases in this work, this end point was found to practically coincide with the end of the fin array: $x_{\text{end,T}} \simeq L_{1} - s_{N}$. 

Later, in Section \ref{sec:msheat_onsetquasi}, we will explain the observed trends for the onset point $x_{\text{periodic,T}}$ from the characteristics of the quasi-developed heat transfer regime, which will be shown to cover the largest part of the development region. 
However, in the next sections, we first discuss the onset of developed macro-scale flow and its implications for the accuracy of our developed Nusselt number correlations. 



\subsection{\label{sec:msheat_onsetdevmacro} Region of developed macro-scale heat transfer}
The heat transfer regime is considered developed from a macro-scale perspective once the macro-scale temperature profiles of the fluid and solid display a linear variation over space, at least at a distance $l_1$ from the side walls, where the spatial moments $\langle \gamma_f \boldsymbol{y} \rangle_{m}^{f}$ and $\langle \gamma_s \boldsymbol{y} \rangle_{m}^{s}$ are zero \cite{buckinx2016macro}: 
\begin{equation}
\begin{aligned}
\langle T \rangle_{m}^{f} &\simeq \langle T_{\text{dev}} \rangle_{m}^{f} \triangleq \mathrm{\nabla{T}} \cdot
\left( \boldsymbol{x} +  \langle \gamma_f \boldsymbol{y} \rangle_{m}^{f}    \right)  + \langle T^{\star} \rangle_{m}^{f} \,, \\
\langle T \rangle_{m}^{s} &\simeq \langle T_{\text{dev}} \rangle_{m}^{s} \triangleq \mathrm{\nabla{T}} \cdot
\left( \boldsymbol{x} +  \langle \gamma_s \boldsymbol{y} \rangle_{m}^{s}    \right)  + \langle T^{\star} \rangle_{m}^{s} \,. 
\end{aligned}
\label{eq:msheat_MSTdev}
\end{equation} 
This follows directly from (\ref{eq:msheat_Tdev}), as the gradient $\mathrm{\nabla{T}}$ is constant and the spatial moments, as well as $\langle T^{\star} \rangle_{m}^{f}$ and $\langle T^{\star} \rangle_{m}^{s}$ are only a function of the transversal coordinate $x_{2}$ in the channel. 
As shown in \cite{buckinx2016macro}, the constant macro-scale temperature gradient is determined here by the incoming heat rate $\langle q_{b} \delta_{bs} \rangle_m = \langle q_{fs} \delta_{fs} \rangle_m $  and the uniform macro-scale flow velocity $\langle \boldsymbol{u}^{\star} \rangle_m$ in the developed region, whose unit direction is $\boldsymbol{e}_{s}$:
\begin{equation}
\begin{aligned}
    \nabla \langle T \rangle_{m} &= \nabla \langle T \rangle_{m}^{f} = \nabla \langle T \rangle_{m}^{s} = \mathrm{\nabla{T}} = \frac{ \langle q_{b} \delta_b \rangle_m }{ \rho_{f} c_{f} \|\langle \boldsymbol{u}^{\star} \rangle_m\| } \boldsymbol{e}_{s} \,.
\end{aligned}
\label{eq:msheat_TemperatureMS_uniform}
\end{equation}

Strictly speaking, the linear temperature profiles (\ref{eq:msheat_MSTdev}) with the constant gradient (\ref{eq:msheat_TemperatureMS_uniform}) will occur at a distance $l_{1}$ after the onset of the periodically developed heat transfer, due to our choice of filter window: $\Omega_{\text{unit}}^{2 \times 2}$. 
Therefore, the proper definition of the region of developed macro-scale heat transfer is given by $\boldsymbol{x} \in \Omega_{\text{dev,T}} \leftrightarrow x_{1} \in \left( x_{\text{dev,T}}, x_{\text{end,T}} - l_{1} \right)$, where the onset point equals $x_{\text{dev,T}} \triangleq x_{\text{periodic,T}} + l_{1}$. 

\begin{figure}[ht!]
\centering
\includegraphics[scale = 0.75]{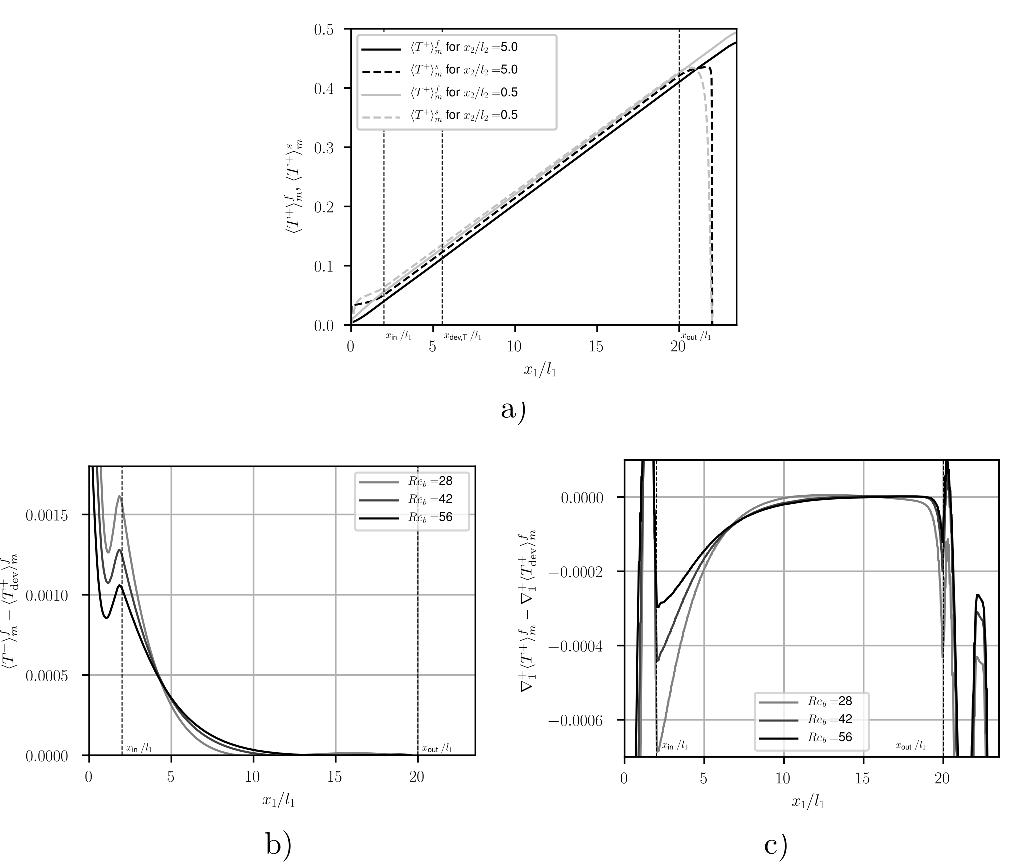}
\caption{\label{fig:msheat_mstemp} Intrinsic fluid and solid macro-scale temperature outside (black) and inside (grey) the side-wall region, when $Pr_{f}=7$, $Re_{b}=28$, $h/l=0.12$, $s/l=0.48$, $t/l=0.02$, $s_{0}=l_{1}$, $s_{N}=2.5 l_{1}$, $N_{1}=20$, and $N_{2}=10$ (a). Intrinsic fluid macro-scale temperature (b) and its gradient (c) along the channel centerline ($x_{2} = L_{2}/2$), when $Pr_{f}=7$, $h/l=0.12$, $s/l=0.48$, $t/l=0.02$, $s_{0}=l_{1}$, $s_{N}=2.5 l_{1}$, $N_{1}=20$, and $N_{2}=10$. Note that the non-dimensional gradient along the main flow direction is noted by $\nabla^{+}_{1} \triangleq \boldsymbol{l}_{1} \cdot \nabla$}
\end{figure}

Yet, we found that the former definition is very conservative because the temperature profiles (\ref{eq:msheat_MSTdev}) appear to be reasonable approximations over a significantly larger region in the channel than just $\Omega_{\text{dev,T}}$. 
This is illustrated in Figure \ref{fig:msheat_mstemp}(a), where the non-dimensional temperature $T^{+}$ is defined as $T^{+} \triangleq (T - T_{\text{in}}) k_f / ( q_{b} l )$. 
In this figure, we observe that even in the region where the macro-scale heat transfer regime is still developing, i.e. $\Omega_{\text{predev,T}}$ where $x_{1} \in (x_{\text{in}},x_{\text{dev,T}})$, the macro-scale temperature fields deviate less than 5\% from the developed linear profiles (\ref{eq:msheat_MSTdev}). 

According to Figure \ref{fig:msheat_mstemp}(b) and Figure \ref{fig:msheat_mstemp}(c), this observation appears to be quite generally valid over a substantial range of Reynolds numbers. 
In particular, all our DNS results indicate that the deviations from the developed temperature profiles remain lower than 20\% of the temperature difference over a single unit cell $| \mathrm{\nabla{T}} \cdot \boldsymbol{l}_{1} |$ in the region behind the first row of fins, up to the last row of the fins. 
On that account, we conclude that, from a practical perspective, the region of developed macro-scale heat transfer coincides with the region over which the weighted porosity $\epsilon_{fm}$ is independent of the streamwise coordinate $x_{1}$, i.e. $x_{1} \in \left( x_{\text{in}},x_{\text{out}} \right)$ with $x_{\text{in}} = s_{0} + l_{1}$ and $x_{\text{out}} = L_{1} - \left( s_{N} + l_{1} \right)$. 
Outside the former region, hence immediately near the channel inlet and outlet, the temperature field does become affected by the specific boundary conditions, as well as the presence of strong porosity gradients. 


\subsection{\label{sec:msheat_accuracy}Accuracy of the developed Nusselt number correlation}

The onset point of developed macro-scale heat transfer can also be identified as the streamwise position after which the local macro-scale heat transfer coefficients $h_{fs}(\boldsymbol{x})$ and $h_{b}(\boldsymbol{x})$ both become spatially constant. 
In particular, $h_{b}(\boldsymbol{x})$ becomes equal to the developed heat transfer coefficient $h_{\text{unit}}$ in the center of the channel \cite{buckinx2016macro,buckinx2017macro}:
\begin{equation}
\label{eq: nusselt number developed correlations}
h_{b}(\boldsymbol{x}) \simeq h_{\text{unit}} \,.
\end{equation}

The developed heat transfer coefficient $h_{\text{unit}}$ 
is obtained from the developed Nusselt number:
$h_{\text{unit}} \triangleq Nu_{\text{unit}} \frac{k_{f}}{l^{2}}$. 
For air ($Pr_f=0.7$ and $k_{s}/k_{f}=10^4$) \cite{vangeffelen2022nusselt}, $Nu_{\text{unit}}$ is given by the correlation 
\begin{equation}
    Nu_{\text{unit}} = c_{0} + c_{1} Re_l, 
\nonumber
\end{equation}
with
\begin{equation}
\begin{aligned}
  c_{0} &= 6.44 (h/l)^{-2} + 9.60(h/l)^{-1.24} + 24.4 (s/l)^{-1.85}, \\
  c_{1} &= 0.112 (s/l-t/l)^{-0.61} (h/l)^{-0.48}\,.\\
\end{aligned}
\label{eq:nusselt_fit_alignedPr07}
\end{equation}
On the other hand, for water ($Pr_f=7$ and $k_{s}/k_{f}=500$),  it is given by
\begin{equation}
    Nu_{\text{unit}} = d_{0} + d_{1} Re_l, 
\nonumber
\end{equation}
with
\begin{equation}
\begin{aligned}
  d_{0} &= 3.84 (h/l)^{-2} + 19.2(h/l)^{-1.39} + 22.3 (s/l)^{-1.87}, \\
  d_{1} &= 1.26 (s/l-t/l)^{-1.07} (t/l)^{0.54} (h/l)^{-0.56} \,.\\
\end{aligned}
\label{eq:nusselt_fit_alignedPr7}
\end{equation}
We remark that correlations (\ref{eq:nusselt_fit_alignedPr07}) and (\ref{eq:nusselt_fit_alignedPr7}) have an average error below 3\% and 4\%, respectively \cite{vangeffelen2022nusselt}. 
Their maximum error is limited to 12\% and 18\%, respectively. 

\begin{figure}[ht!]
\centering
\includegraphics[scale = 1.00]{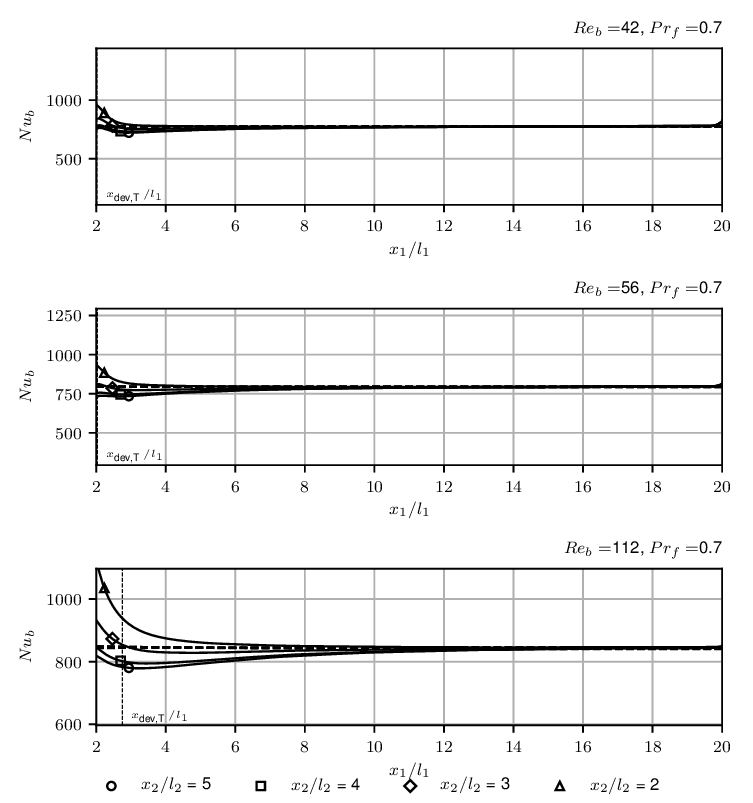}
\caption{\label{fig:msheat_Nu_x_t0.02_h0.12_s0.48_1} Macro-scale heat transfer coefficient $Nu_{b}$ (full) and its prediction $Nu_{\text{unit}}$ (dashed) by the developed correlations from \cite{vangeffelen2022nusselt}, when $Pr_f=0.7$, $h/l=0.12$, $s/l=0.48$, $t/l=0.02$, $s_{0}=l_{1}$, $s_{N}=2.5 l_{1}$, $N_{1}=20$, and $N_{2}=10$}
\end{figure}

\begin{figure}[ht!]
\centering
\includegraphics[scale = 1.00]{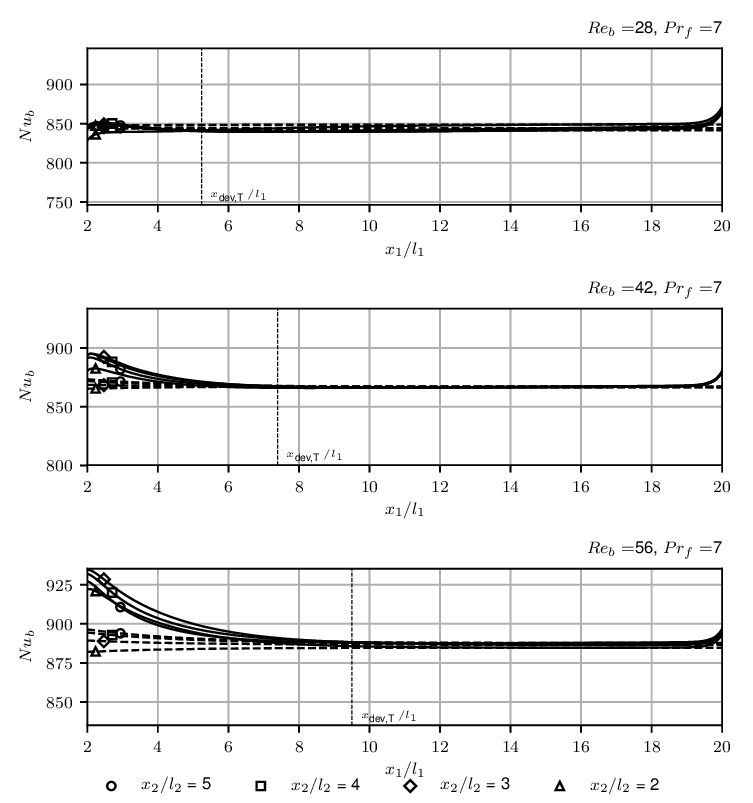}
\caption{\label{fig:msheat_Nu_x_t0.02_h0.12_s0.48_2} Macro-scale heat transfer coefficient $Nu_{b}$ (full) and its prediction $Nu_{\text{unit}}$ (dashed) by the developed correlations from \cite{vangeffelen2022nusselt}, when $Pr_f=7$, $h/l=0.12$, $s/l=0.48$, $t/l=0.02$, $s_{0}=l_{1}$, $s_{N}=2.5 l_{1}$, $N_{1}=20$, and $N_{2}=10$}
\end{figure}

In principle, the criterion $h_b \simeq h_{\text{unit}}$ will result in a different value for the onset point of the developed regime (\ref{eq: nusselt number developed correlations}) in comparison with the former criterion (\ref{eq:msheat_MSTdev}). 
However, according to our numerical results, both criteria are practically identical, as Figures \ref{fig:msheat_Nu_x_t0.02_h0.12_s0.48_1} and \ref{fig:msheat_Nu_x_t0.02_h0.12_s0.48_2} reveal. 

In Figures \ref{fig:msheat_Nu_x_t0.02_h0.12_s0.48_1} and \ref{fig:msheat_Nu_x_t0.02_h0.12_s0.48_2}, we show the actual macro-scale heat transfer coefficient $h_{b}(\boldsymbol{x})$ along a channel at four different lateral positions with an array of $20 \times 10$ offset strip fin unit cells. 
In this figure, we have represented $h_{b}(\boldsymbol{x})$ by the local Nusselt number $Nu_{\text{b}}(\boldsymbol{x}) \triangleq h_b l^2/k_f$. 
Figure \ref{fig:msheat_Nu_x_t0.02_h0.12_s0.48_1} corresponds to an air flow, while Figure \ref{fig:msheat_Nu_x_t0.02_h0.12_s0.48_2} corresponds to a water flow. 
The considered Reynolds numbers for both fluids are relatively low to moderate ($Re_b=28-112$), so that the resulting Péclet numbers are low to moderate for air ($Pe \simeq 30-80$) and moderate for water ($Pe \simeq 200-400$), as of interest for many micro- and mini-channel applications \cite{hachemi1999experimental,hong2009three,do2016experimental}. 

Figures \ref{fig:msheat_Nu_x_t0.02_h0.12_s0.48_1} and \ref{fig:msheat_Nu_x_t0.02_h0.12_s0.48_2} illustrate that the actual macro-scale heat transfer coefficient $h_{b}$ deviates less than 5\% from its constant value for $x_1 \geq x_{\text{periodic,T}} = x_{\text{dev,T}}-l_{1}$, when viewed along the centerline $x_2=5 l_2$ of the channel. 
This demonstrates that the onset point $x_{\text{periodic,T}}$ is also a good measure to indicate where the developed Nusselt number correlation becomes valid, even though $x_{\text{periodic,T}}$ is based on a criterion for the underlying temperature field instead (\ref{eq:msheat_Tdev}). 
Due to numerical errors, however, the constant value of $h_{b}$ in the developed value region still differs slightly from the constant value $h_{\text{unit}}$ predicted by the correlations (\ref{eq:nusselt_fit_alignedPr07}) and (\ref{eq:nusselt_fit_alignedPr7}): the difference $| h_{b} - h_{\text{unit}} |/h_{b}$ may be up to 7\%. 
These numerical errors originate mainly from the limited accuracy of the correlation, as the discretization error on $h_{b}$ is around 1\%. 

Figures \ref{fig:msheat_Nu_x_t0.02_h0.12_s0.48_1} and \ref{fig:msheat_Nu_x_t0.02_h0.12_s0.48_2} also give an impression of the typical accuracy of the developed Nusselt number correlations in the region where the flow and temperature fields are developing, at least for Péclet numbers in relevant micro- and mini-channel applications. 
In the figure, we have compared the actual local macro-scale heat transfer coefficient $h_{b}(\boldsymbol{x})$ with its local prediction $h_{\text{unit}}(\boldsymbol{x})$ based on the local Nusselt number $Nu_{\text{unit}}\left( \boldsymbol{x} \right)$. 
Hereto, we have evaluated $Nu_{\text{unit}}\left( \boldsymbol{x} \right)$ based on the local Reynolds number $Re_{l} \left( \boldsymbol{x} \right) \triangleq \rho_{f} \|\langle \boldsymbol{u} \rangle_{m} \| l / \mu_{f}$, for the given Prandtl number and unit cell geometry. 
Essentially, this means that we treat the macro-scale heat transfer regime locally as almost developed, so that the local macro-scale heat transfer coefficient is governed by the periodically developed temperature equations from \cite{buckinx2017macro}, even when the flow is still developing. 
The main takeaway from this figure is that the local value $h_{\text{unit}}\left( \boldsymbol{x} \right)$ predicted by the developed Nusselt number correlation is also quite accurate over a significant part of the developing flow and heat transfer region. 
After all, even upstream of the strict onset point $x_{\text{dev,T}}$, the developed Nusselt number correlations (\ref{eq:nusselt_fit_alignedPr07})-(\ref{eq:nusselt_fit_alignedPr7}) predict the actual macro-scale heat transfer coefficient $h_{b}$ with a mean and maximum relative error $| h_{b} - h_{\text{unit}} |/h_{b}$ of 10\% and 25\%, respectively. 
One part of the explanation is that the actual macro-scale heat transfer coefficient in the developing flow and heat transfer region only deviates between 5\% and 30\% from its constant value in the developed region. 
However, the small error between the actual heat transfer coefficient $h_{b}(\boldsymbol{x})$ and its local approximation $h_{\text{unit}}\left( \boldsymbol{x} \right)$ immediately upstream of the point $x_{\text{dev,T}}$ can also be explained in a different way. 
As we will show in the next section, this error is determined by the inherent characteristics of the quasi-periodically developed heat transfer regime in $\Omega_{\text{predev,T}}$. 

Given the large range of Péclet numbers covered, we expect that the former results are indicative of a wide range of practical applications, especially those where the channel height-to-length ratio $h/l$ is below 0.2 and the Reynolds number remains below 150 \cite{vangeffelen2022nusselt}. 
After all, for such geometries, the Péclet number has the most significant effect on the onset of developed heat transfer, as shown in Section \ref{sec:msheat_onsetperiodic}. 

The most severe limitation of the results in Figures \ref{fig:msheat_Nu_x_t0.02_h0.12_s0.48_1} and \ref{fig:msheat_Nu_x_t0.02_h0.12_s0.48_2} is the assumed parabolic inlet velocity profile for the flow simulations \cite{vangeffelen2023developed}. 
For strongly asymmetric inlet flow, the former results will no longer be representative. 
However, in practical applications, channel manifolds are designed to establish ideal operating conditions, effectively ensuring a more uniform inlet velocity profile. 
Such more uniform profiles induce smaller velocity perturbations near the channel inlet than numerically predicted, according to the experimental measurements of Eneren et al. \cite{eneren2024flow}. 
In turn, these smaller velocity perturbations imply that also the deviations from the developed heat transfer regime will be even smaller in practice, across different channel geometries. 
Together with the broad Péclet number range considered, this reinforces the generality of our conclusions, supporting the applicability of our findings across various micro- and mini-channel configurations with offset strip fins. 

For the former reasons, we thus expect that our developed heat transfer correlations can adequately model the macro-scale heat transfer regime in the most commonly applied micro- and mini-channels with offset strip fins. 
Certainly, for the conditions studied here, the effects of developing heat transfer are negligible at the macro-scale level when an accuracy level of 10\% to 20\% for the heat transfer coefficient is deemed acceptable. 
In that case, the modelling errors due to the use of developed Nusselt number correlations are larger than the actual deviations from the developed heat transfer coefficient in the developing region.

\newpage
\clearpage


\newpage
\clearpage

\section{\label{sec:msheat_onsetquasi}Onset of quasi-developed macro-scale heat transfer}
The preceding numerical simulations have indicated that the thermal development lengths in offset strip fin micro- and mini-channels are notably short. 
Furthermore, they revealed that deviations from the established macro-scale temperature profiles are small, and they exposed a good accuracy of developed heat transfer correlations even upstream of the developed region. 
In this section, we will show that these three main observations can be attributed to the characteristics of the quasi-developed heat transfer regime, as recently described by \cite{buckinx2024arxiv}. 
According to our next analysis, this regime prevails over almost the entire region $\Omega_{\text{predev,T}}$ where the (macro-scale) temperature field is still developing.

\subsection{\label{sec:msheat_onsetquasiperiodic} Onset of the quasi-periodically developed heat transfer regime}


In the quasi-periodically developed heat transfer regime, the temperature field $T$ converges asymptotically towards the developed temperature field $T_{\text{dev}}$ from (\ref{eq:msheat_Tdev}) along the main flow direction via a single exponential mode \cite{buckinx2024arxiv}: 
\begin{equation}
    T \simeq T_{\text{dev}} + \Theta \exp \left( - \lambda_{T} x_{1}  \right). 
    \label{eq:msheat_quasitemp}
\end{equation}
Here, the mode's eigenvalue $\lambda_{T}$ is a constant, while the mode's amplitude $\Theta $ is spatially periodic along the streamwise coordinate $x_{1}$: $\Theta \left( \boldsymbol{x} + \boldsymbol{l}_{1} \right) = \Theta \left( \boldsymbol{x} \right)$. 
We therefore define the onset point of quasi-periodically developed heat transfer, $x_{\text{quasi-periodic,T}}$, as the streamwise coordinate $x_{1}$ after which the temperature field agrees with expression (\ref{eq:msheat_quasitemp}) within 10\% of the temperature difference over a unit cell in the developed region, so $| T - T_{\text{dev}}  -\Theta \exp \left( - \lambda_{T} x_1 \right) | \leq 0.1 | \mathrm{\nabla{T}} \boldsymbol{\cdot} \boldsymbol{l}_1| $ for $ x_1 \geq  x_{\text{quasi-periodic,T}}$. 
The former definition of $x_{\text{quasi-periodic,T}}$ is analogous to that of $x_{\text{periodic,T}}$. 
The extent of the quasi-periodically developed heat transfer region is thus given by $x_{1} \in (x_{\text{quasi-periodic,T}},x_{\text{periodic,T}})$. 

\begin{figure}[ht]
\centering
\includegraphics[scale = 0.75]{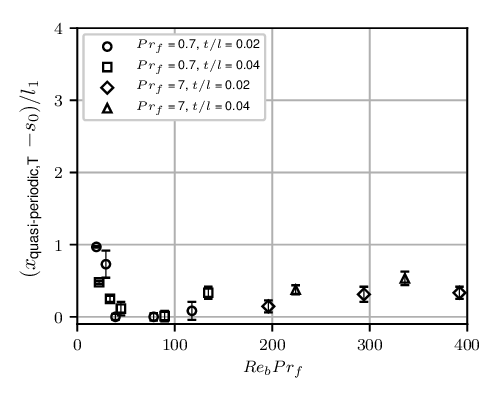}
\caption{\label{fig:onsetT_PE_t} Influence of the Péclet number on the onset of the quasi-developed heat transfer, when $N_{1}=20$, $h/l=0.12$, $s/l=0.48$}
\end{figure}

In Figure \ref{fig:onsetT_PE_t}, the onset point $x_{\text{quasi-periodic,T}}$ has been quantified for two channel geometries and two Prandtl numbers. 
This figure illustrates our more general observation that for all the cases considered in this work, the onset point of the quasi-periodically developed heat transfer regime virtually coincides with the first row of the fin array. 
In other words, in these kinds of micro- and mini-channels, the temperature field throughout the entire array of offset strip fins can be regarded as quasi-developed, just like the flow field \cite{vangeffelen2023developed}. 
Nevertheless, it is important to note that the onset point $x_{\text{quasi-periodic,T}}$ in principle depends on the Prandtl number of the fluid. 
So, for higher Prandtl numbers than those considered in this work, one may expect the onset point $x_{\text{quasi-periodic,T}}$ to be located (slightly) more downstream than the onset point of quasi-periodically developed flow $x_{\text{quasi-periodic}}$. 
For the low to moderate Prandtl numbers in this study, however, this spatial lagging between $x_{\text{quasi-periodic,T}}$ and $x_{\text{quasi-periodic}}$ is not visible. 

\subsection{\label{sec:msheat_eigenvaluesquasiperiodic}Eigenvalues and perturbations for quasi-periodically developed heat transfer}

Since the temperature field can be practically described as quasi-periodically developed over the entire offset strip fin array, the onset and extent of the periodically developed heat transfer regime are, in the first place, determined by the eigenvalue of the exponentially decaying mode (\ref{eq:msheat_quasitemp}). 
In the second place, they are also affected by the peak value of the mode's amplitude, which results from the specific inlet conditions. 
The former statements can be understood from the relation between the onset point of developed heat transfer $x_{\text{periodic,T}}$, the eigenvalue $\lambda_{T}$ and mode amplitude $\Theta$, which follows directly from equation (\ref{eq:msheat_quasitemp}):
\begin{equation}
x_{\text{periodic,T}} \simeq \frac{1}{\lambda_{T}} \ln \left( \frac{\epsilon_{0,T}}{\epsilon_{T}} \right). 
\label{eq:msheat_lambda_onset}
\end{equation}
In this relation, $\epsilon_{T}$ is the criterion used for defining the onset point $x_{\text{periodic,T}}$. 
More specifically, $\epsilon_{T}$ is defined by $ | T - T_{\text{dev}} | / | \mathrm{\nabla{T}} \cdot \boldsymbol{l}_{1} | \leqslant \epsilon_{T}$ for $x_{1} \geqslant x_{\text{periodic,T}}$. 
As explained in Section \ref{sec:msheat_onsetperiodic}, we have adopted the criterion $\epsilon_{T}=0.1$ in this work. 
Further, the relative temperature difference $\epsilon_{0,T}$ is a measure for the maximum deviation from the developed temperature profile at the onset point of the quasi-periodically developed heat transfer regime: $ | T - T_{\text{dev}} | / | \mathrm{\nabla{T}} \cdot \boldsymbol{l}_{1} | \leqslant \epsilon_{0,T} \exp \left( - \lambda_{T} x_{\text{quasi-periodic,T}} \right)$ for $x_{1} \geqslant x_{\text{quasi-periodic,T}}$. 
Therefore, $\epsilon_{0,T}$ characterizes the peak value of the mode amplitude at $x_{\text{quasi-periodic,T}}$: 
$\epsilon_{0,T} = \max \limits_{\boldsymbol{x} \in S} \left| \Theta / ( \mathrm{\nabla{T}} \cdot \boldsymbol{l}_{1} ) \right| $ with $S \triangleq \{\boldsymbol{x} \vert \, x_1=x_{\text{quasi-periodic,T}} \}$. 
The exact value of this factor $\epsilon_{0,T}$, also called the \textit{temperature perturbation size}, depends on how the temperature field becomes developed and, as such, the specific inlet conditions and channel inlet geometry. 
For instance, $\epsilon_{0,T}$ is influenced by the distance $s_0$ between the inlet and the fin array, although it is not affected by the outlet geometry nor $s_N$. 

From relationship (\ref{eq:msheat_lambda_onset}), we learn that the relatively short thermal development lengths in offset strip fin micro- and mini-channels can be understood as a consequence of the relatively large eigenvalues $\lambda_T$ of the temperature modes in these channels. 
Because the eigenvalues $\lambda_T$ are the solutions to a periodic eigenvalue problem on a single row of the entire fin array, which is given in \cite{buckinx2024arxiv}, their relatively large magnitude is an inherent property of the geometry of the offset strip fin row. 

From relationship (\ref{eq:msheat_lambda_onset}), we also learn that the onset point of periodically developed heat transfer will scale as $x_{\text{periodic,T}} \sim 1/\lambda_{T}$ when the peak value of the mode amplitude has no significant effect on the logarithmic factor $\ln \left( \epsilon_{0,T} / \epsilon_{T} \right)$. 
As we will demonstrate next, the scaling law $x_{\text{periodic,T}} \sim 1/\lambda_{T}$ explains well the influence of the Reynolds number, the Prandtl number and channel aspect ratio on the onset point $x_{\text{periodic,T}}$. 

\begin{figure}[ht!]
\centering
\begin{minipage}{.475\textwidth}
\hspace{-8mm}
\includegraphics[scale = 0.75]{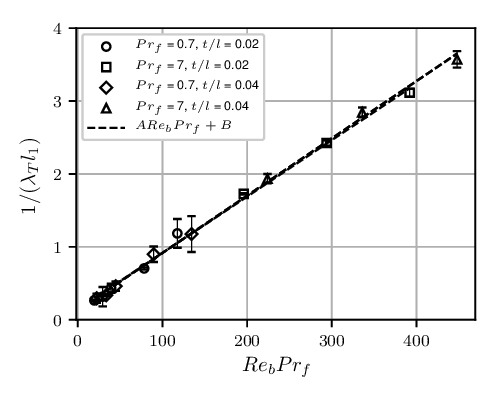}
\caption{\label{fig:msheat_lambdaT_PE_t} Influence of the Péclet number on the eigenvalue of the quasi-developed heat transfer, when $N_{1}=20$, $h/l=0.12$, $s/l=0.48$ \newline}
\end{minipage}
\hfill
\begin{minipage}{.475\textwidth}
\hspace{-5mm}
\includegraphics[scale = 0.75]{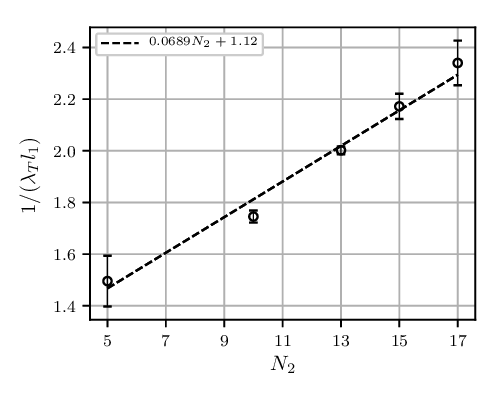}
\caption{\label{fig:msheat_lambdaT_N2} Influence of the channel aspect ratio on the eigenvalue of the quasi-developed heat transfer, when $Pr_{f}=7$, $Re_{b}=28$, $N_{1}=20$, $h/l=0.12$, $s/l=0.48$, $t/l=0.02$}
\end{minipage}
\end{figure}

The fact that the scaling law $x_{\text{periodic,T}} \sim 1/\lambda_{T}$ explains the previously observed linear relation between the onset point $x_{\text{periodic,T}}$ and the Reynolds number $Re_{b}$ appears from the linear relation between $1/\lambda_{T}$ and $Re_{b}$ in Figure \ref{fig:msheat_lambdaT_PE_t}. 
In this Figure, the linear variation of the inverse eigenvalue $1/\lambda_{T}$ with the Reynolds number $Re_{b}$ is depicted for two channel geometries and two Prandtl numbers. 
The uncertainty bars on the eigenvalues obtained through DNS indicate the possible error margins due to discretization errors and the employed least-square regression technique. 
The notion that $x_{\text{periodic,T}}/l_{1} \sim 1/(\lambda_{T} l_{1}) \sim ARe_{b} + B$ for some constants $A$ and $B$ is also in line with our observation that the logarithmic perturbation size $\ln \left( \epsilon_{0,T} / \epsilon_{T} \right)$ remains constant to a first-order approximation over a wide range of Reynolds numbers, according to our DNS results.
For instance, over the range of Reynolds numbers in Figure \ref{fig:msheat_lambdaT_PE_t}, we found that $\epsilon_{0,T} \simeq 3.67$ when $t/l=0.02$, and $\epsilon_{0,T} \simeq 6.56$ when $t/l=0.04$ with a relative error below 10\%. 
This observation remains valid as long as the shape of the velocity profile and temperature profile at the channel inlet are kept fixed, hence independent of the Reynolds number $Re_{b}$. 
Strong changes in the inlet profiles could of course induce very different temperature perturbations, so that the observed similarity of the velocity and temperature profiles at different Reynolds numbers could get lost. 
We remark that the inversely linear relationship between the temperature eigenvalue $\lambda_{T}$ and Reynolds number $Re_{b}$ expresses a direct analogy with the quasi-periodically developed flow regime, in which the eigenvalue $\lambda$ of the velocity mode is also an inversely linear function of $Re_{b}$ \cite{vangeffelen2023developed}. 

As Figure \ref{fig:msheat_lambdaT_PE_t} illustrates, the influence of the Prandtl number $Pr_f$ on the eigenvalue $\lambda_{T}$ can be captured by an inversely linear correlation in terms of the Péclet number. 
For both Prandtl numbers shown in Figure \ref{fig:msheat_lambdaT_PE_t}, the following correlations hold with a maximum relative error of 10\%: $\lambda_{T} l_{1} \simeq 1 / \left( 0.00812 Pe + 0.0999 \right)$ for $t/l=0.02$, and $\lambda_{T} l_{1} \simeq 1 / \left( 0.00810 Pe + 0.0994 \right)$ for $t/l=0.04$. 
Because the temperature perturbation size $\epsilon_{0,T}$ barely changes with the Péclet number, so that $x_{\text{periodic,T}} \sim 1/\lambda_{T}$, these correlations explain directly why the onset point of the periodically developed heat transfer regime $x_{\text{periodic,T}}$ scales linearly with the Péclet number for a single offset strip fin channel geometry, as we described in Section \ref{sec:msheat_onsetperiodic}. 

When we compare the eigenvalues $\lambda_{T}$ of the temperature modes from Figure \ref{fig:msheat_lambdaT_PE_t} with the eigenvalues $\lambda$ of the velocity modes in the quasi-periodically developed flow regime \cite{vangeffelen2023developed}, we find that $\lambda_{T}$ is a factor of four to eight larger than $\lambda$ when $Pr_{f} = 0.7$. 
This supports our earlier observation that the onset point $x_{\text{periodic,T}}$ is located upstream of the onset point of the periodically developed flow regime $x_{\text{periodic}}$ when $Pr_{f} = 0.7$, since $x_{\text{periodic,T}}/x_{\text{periodic}} \sim \lambda/\lambda_{T}$, because of $x_{\text{quasi-periodic,T}}\simeq x_{\text{quasi-periodic}} \simeq s_0 + l_{1}$. 
On the contrary, when $Pr_{f} = 7$, both $x_{\text{periodic,T}}$ and $x_{\text{periodic}}$ have the same order of magnitude, as $\lambda$ and $\lambda_{T}$ exhibit similar magnitudes as well. 


In Figure \ref{fig:msheat_lambdaT_N2}, we show the dependence of the eigenvalue $\lambda_{T}$ on the number of units cells in the lateral direction $N_{2}$, hence the inverse channel aspect ratio $L_{2}/L_{3}$, for a rather low Reynolds number $Re_b =28$, and a small channel height $h/l=0.12$. 
It can be seen that the inverse eigenvalue $1/\lambda_{T}$ is again a linear function of $N_{2}$, just like the inverse eigenvalue $1/\lambda$ of the velocity modes in quasi-periodically developed flow \cite{vangeffelen2023developed}. 
For the data in Figure \ref{fig:msheat_lambdaT_N2}, we obtained the correlation $\lambda_{T} l_{1} \simeq 1 / \left( 0.0689 N_{2} + 1.12 \right)$, which is accurate within a maximum relative error of 4\%. 
In addition, we found that $\epsilon_{0,T} \simeq 3.33$ within an error of 10\%, so that the factor $\ln \left( \epsilon_{0,T} / \epsilon_{T} \right)$ varies about 3\% with $N_{2}$. 
So, the logarithmic perturbation size $\ln \left( \epsilon_{0,T} / \epsilon_{T} \right)$ is virtually independent of the channel's aspect ratio. 
The scaling law $x_{\text{periodic,T}} \sim 1/\lambda_{T} $ therefore recovers the correct relationship between the onset point $x_{\text{periodic,T}}$ and aspect ratio as described before, i.e. $x_{\text{periodic,T}} \sim 1/\lambda_{T} \sim A N_{2} + B$, where $A$ and $B$ are constants.



\begin{figure}[ht!]
\centering
\begin{minipage}{.475\textwidth}
\hspace{-8mm}
\includegraphics[scale = 0.75]{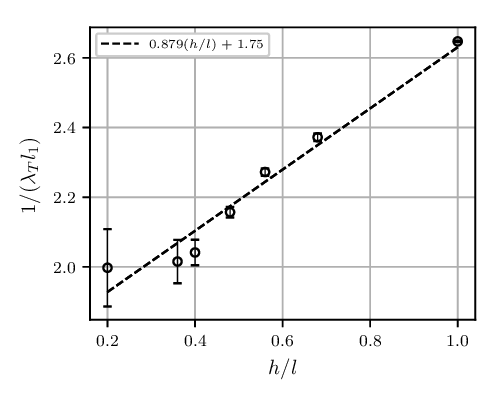}
\caption{\label{fig:msheat_lambdaT_h} Influence of the fin height-to-length ratio on the eigenvalue of the quasi-developed heat transfer, when $Pr_{f}=7$, $Re_{b} l / (2 L_{3}) = \rho_{f} u_{b} l / \mu_{f}=600$, $N_{1}=20$, $N_{2}=10$, $s/l=0.12$, $t/l=0.02$}
\end{minipage}
\hfill
\begin{minipage}{.475\textwidth}
\hspace{-5mm}
\includegraphics[scale = 0.75]{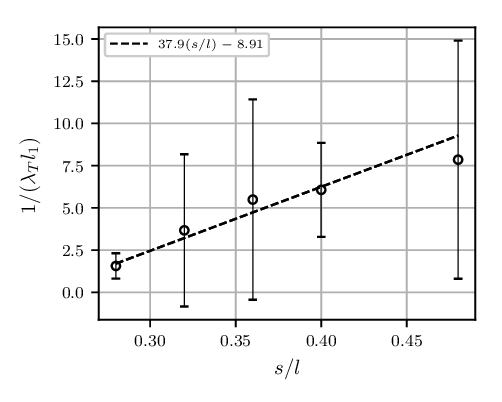}
\caption{\label{fig:msheat_lambdaT_s} Influence of the fin pitch-to-length ratio on the eigenvalue of the quasi-developed heat transfer, when $Pr_{f}=7$, $Re_{b}=192$, $N_{1}=20$, $N_{2}=10$, $h/l=0.12$, $t/l=0.04$}
\end{minipage}
\end{figure}

Whereas the scaling law $x_{\text{periodic,T}} \sim 1/\lambda_{T}$ accounts for the observed dependence of the onset point $x_{\text{periodic,T}}$ on the Péclet number and the aspect ratio, it falls short of describing the influence of the fin height-to-length ratio $h/l$ and fin pitch-to-length ratio $s/l$. 
The reason is that the latter parameters significantly affect the temperature perturbation size. 
According to Figure \ref{fig:msheat_lambdaT_h}, the dependence of the eigenvalue $\lambda_{T}$ on the fin height-to-length ratio $h/l$ is described by an inversely linear relationship: $\lambda_{T} l_{1} \simeq 1 / \left( 0.879 (h/l) + 1.75 \right)$ with an accuracy of 4\%. 
This agrees with our expectation that the channel height $h$ and channel width $L_2$ have a similar impact on the transport of thermal energy. 
Yet, for the data in Figure \ref{fig:msheat_lambdaT_h}, we found that the temperature perturbation size satisfies $\epsilon_{0,T} \simeq 3.55 (h/l) - 0.156$ within an error of 10\%, so that the factor $\ln \left( \epsilon_{0,T} / \epsilon_{T} \right)$ almost doubles over the interval $h/l \in (0.2,1.0)$. 

Similarly, the fin pitch-to-length ratio $s/l$ has a strong influence on both the eigenvalue $\lambda_{T}$ and induced temperature perturbation size $\epsilon_{0,T}$. 
Figure \ref{fig:msheat_lambdaT_s} illustrates this for a water flow ($Pr_{f}=7$) at a moderate Reynolds number $Re_{b}=192$ through a compact channel with a dense array of offset strip fins: $N_{2}=10$, $h/l=0.12$, $t/l=0.04$ and $s/l \in (0.28,0.48)$. 
It can be seen that the eigenvalue $\lambda_{T}$ in this case varies with the fin pitch-to-length ratio $s/l$ according to $\lambda_{T} l_{1} \simeq 1 / \left( 37.9 (s/l) - 8.91 \right)$, within a relative error of 5\%. 
For the same DNS data, it was found that within 10\%, the temperature perturbation size is given by $\epsilon_{0,T} \simeq 19.8 (s/l) - 4.47$. 
Therefore, the observed scaling of $x_{\text{periodic,T}}$ with $s/l$ in Section \ref{sec:msheat_onsetperiodic} is a complex interplay between the actual inlet conditions and the eigenmodes in the quasi-periodically developed heat transfer regime. 
The strong effect of the fin height-to-length ratio and pitch-to-length ratio on the temperature perturbations can be understood from their effect on the flow field.
When the fin height or pitch increases, the flow will experience relatively less friction due to the channel walls or fin sides.
In turn, perturbations in the temperature field will be more strongly advected in the streamwise direction with less dampening due to transversal diffusion since the transversal distance to the channel walls or fin sides is larger.

\begin{figure}[ht!]
\centering
\includegraphics[scale = 0.75]{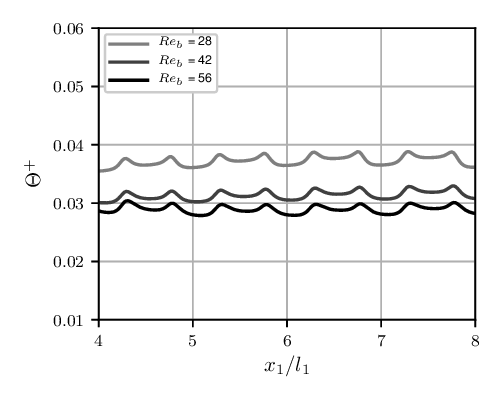}
\caption{\label{fig:Tmode_RE} Temperature mode amplitude along the channel centerline ($x_{2} = L_{2}/2$, $x_{3} = L_{3}/2$), when $Pr_{f}=7$, $h/l=0.12$, $s/l=0.48$, $t/l=0.02$, $s_{0}=l_{1}$, $s_{N}=2.5 l_{1}$, $N_{1}=20$, and $N_{2}=10$}
\end{figure}

Finally, we illustrate the temperature modes in a typical channel geometry for three rather low Reynolds numbers in Figure \ref{fig:Tmode_RE}. 
In this figure, the non-dimensional temperature mode is defined as $\Theta^{+} \triangleq \Theta k_f / ( q_{b} l )$. 
It can be recognized that the maximum of the dimensionless mode amplitude $\Theta^{+}$ is inversely proportional to Reynolds number $Re_{b}$. 
This means that the maximum of the actual mode amplitude $\Theta$ varies inversely with the mass flow rate, and hence the developed volume-averaged velocity $\|\langle \boldsymbol{u}^{\star} \rangle\|$. 
This, in turn, implies that the maximum of the temperature mode $\Theta$ is proportional to the temperature difference over a unit cell $| \mathrm{\nabla{T}} \cdot \boldsymbol{l}_{1} |$, so that the temperature perturbation size $\epsilon_{0,T}$ can be treated as independent of the flow rate and Reynolds number, as discussed before. 

In conclusion, our observations show that the quick onset of the periodically developed heat transfer regime is mainly explained by the large magnitude of the mode eigenvalues and their scaling laws with respect to the Reynolds number, Péclet number, and aspect ratio.
The mode amplitudes are of secondary importance.
They account for some of the effects of the unit cell geometry and, thus, the array porosity on the development length. 







\subsection{\label{sec:msheat_onsetquasidevmacro} Region of quasi-developed macro-scale heat transfer}

Given that the heat transfer regime can be regarded as quasi-periodically developed nearly from the start of the offset strip fin array in micro- and mini-channels, also the macro-scale temperature field can be characterized as quasi-developed across most of the fin array. 
The small deviations from the developed macro-scale temperature profile discussed in Section \ref{sec:msheat_onsetdevmacro} are thus primarily explained by the fast decaying temperature modes in $\Omega_{\text{predev,T}}$, since the eigenvalue $\lambda_T$ is quite large:
\begin{equation}
\begin{aligned}
\langle T \rangle_{m}^{f} &\simeq \mathrm{\nabla{T}} \cdot \left( \boldsymbol{x} + \langle \gamma_f \boldsymbol{y} \rangle_{m}^{f} \right)  + \langle T^{\star} \rangle_{m}^{f} + \langle \Theta \rangle_{m}^{f} \exp \left( - \lambda_{T} x_{1}  \right), \\
\langle T \rangle_{m}^{s} &\simeq \mathrm{\nabla{T}} \cdot \left( \boldsymbol{x} + \langle \gamma_s \boldsymbol{y} \rangle_{m}^{s} \right) + \langle T^{\star} \rangle_{m}^{s} + \langle \Theta \rangle_{m}^{s} \exp \left( - \lambda_{T} x_{1}  \right).
\label{eq:msheat_quasitempMS}
\end{aligned}
\end{equation}
Strictly speaking, the region of the quasi-developed macro-scale heat transfer regime, where the former temperature profiles (\ref{eq:msheat_quasitempMS}) apply, is given by $x_{1} \in (x_{\text{quasi-dev,T}},x_{\text{dev}})$, where the onset point corresponds to $x_{\text{quasi-dev,T}} = x_{\text{quasi-periodic,T}} + l_{1}$. 
Nevertheless, in practise we may take $x_{\text{quasi-dev,T}} \simeq x_{\text{quasi-periodic,T}} $ for the same reasons as why we can take $x_{\text{dev,T}} \simeq x_{\text{periodic,T}}$. 

\begin{figure}[ht!]
\centering
\begin{minipage}{.475\textwidth}
\hspace{-5mm}
\includegraphics[scale = 0.75]{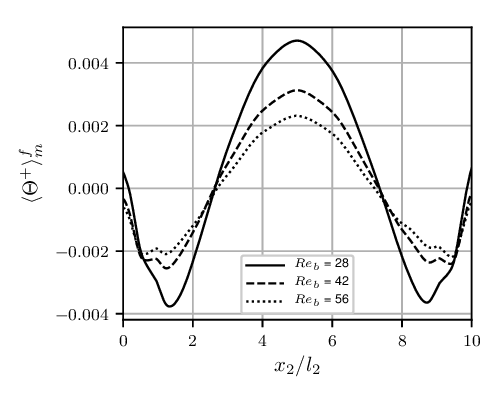}
\caption{\label{fig:msheat_Thetaf_profile_PE} Influence of the Reynolds number on the intrinsic macro-scale fluid temperature mode for quasi-developed heat transfer, when $Pr_{f}=7$, $N_{2}=10$, $h/l=0.12$, $s/l=0.48$, $t/l=0.02$}
\end{minipage}
\hfill
\begin{minipage}{.475\textwidth}
\hspace{-5mm}
\includegraphics[scale = 0.75]{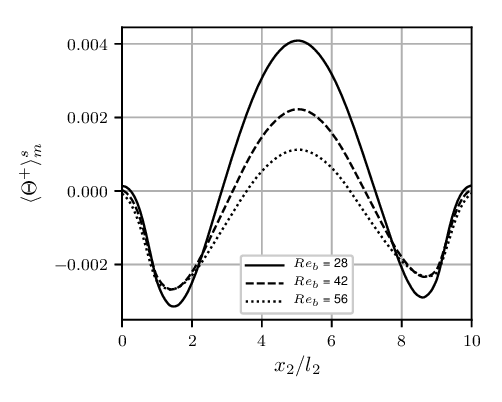}
\caption{\label{fig:msheat_Thetas_profile_PE} Influence of the Reynolds number on the intrinsic macro-scale solid temperature mode for quasi-developed heat transfer, when $Pr_{f}=7$, $N_{2}=10$, $h/l=0.12$, $s/l=0.48$, $t/l=0.02$}
\end{minipage}
\end{figure}

The macro-scale temperature mode amplitudes $\langle \Theta \rangle_{m}^{f}$ and $\langle \Theta \rangle_{m}^{s}$, in the former expressions (\ref{eq:msheat_quasitempMS}) for the intrinsic macro-scale temperatures are only a function of the transversal coordinate $x_{2}$ in the channel. 
In Figures \ref{fig:msheat_Thetaf_profile_PE} and \ref{fig:msheat_Thetas_profile_PE}, they are illustrated in their dimensionless form for the same Reynolds number, Prandtl number and geometry as in Figure \ref{fig:Tmode_RE}. 
Their W-like shape strongly resembles the shape of the velocity modes $\langle \boldsymbol{U} \rangle_m^f$ in the quasi-developed flow region because the temperature modes are advected by the same periodic flow field $\boldsymbol{u}^{\star}$ as the velocity modes, according to the eigenvalue problem from \cite{buckinx2022arxiv}. 
This shape reflects that the deviations from the developed macro-scale temperature profile are most pronounced at the center of the channel and at a distance of $l_{1}$ from the side walls. 

Theoretically, the magnitude (or peak value) of both amplitudes $\langle \Theta \rangle_{m}^{f}$ and $\langle \Theta \rangle_{m}^{s}$ depends on the specific inlet conditions and channel inlet geometry. 
Nevertheless, it can be seen from Figures \ref{fig:msheat_Thetaf_profile_PE} and \ref{fig:msheat_Thetas_profile_PE} that the influence of the Reynolds number $Re_{b}$ on the shape of the macro-scale temperature modes is small. 
As such, their profile can be assumed to be universal for a single channel geometry. 
For example, for the macro-scale temperature modes in Figures \ref{fig:msheat_Thetaf_profile_PE} and \ref{fig:msheat_Thetas_profile_PE}, we can define the following reference profile to characterize their shape at a distance $l_2$ away from the side walls:
\begin{equation}
\label{eq: macro-scale temperature reference mode}
\Theta_{\text{ref}} \triangleq \frac{q_{b} l}{k_{f}} \sin \left( \frac{2 \pi}{L_{2}} (x_2 - L_{2}/4) \right). 
\end{equation}
Closer to the side walls, however, the actual profile diverges from this reference profile due to the imposed no-heat-flux boundary condition and the loss of transversal periodicity in the solid material distribution $\gamma_s$. 
Based on the reference profile (\ref{eq: macro-scale temperature reference mode}), the modes for all Reynolds numbers in Figures \ref{fig:msheat_Thetaf_profile_PE} and \ref{fig:msheat_Thetas_profile_PE} can be correlated as  $\langle \Theta \rangle_{m}^{f} / \Theta_{\text{ref}} \simeq 0.940 Re_{b}^{-1} - 7.82e\text{-}5$ and $\langle \Theta \rangle_{m}^{s} / \Theta_{\text{ref}} \simeq 1.15 Re_{b}^{-1} - 0.00177$, at least within a maximum relative error of 10\%. 
The small coefficients in these correlations verify again our observation that the thermal perturbation size $\epsilon_{0,T}$ is approximately independent of $Re_{b}$, as discussed in Section \ref{sec:msheat_eigenvaluesquasiperiodic}. 


\subsection{\label{sec:msheat_closure}Closure for the quasi-developed heat transfer region}


The relatively large eigenvalues and small induced magnitudes of the macro-scale temperature modes in the quasi-developed heat transfer region directly account for the good accuracy of the developed Nusselt Number correlations even where the temperature field is still developing. 
Mathematically, this can also be understood from the asymptotically correct forms of the true macro-scale heat transfer coefficients $h_{fs}$ and $h_b$ in $\Omega_{\text{predev,T}}$, which are given by 
\begin{equation}
h_{fs} \simeq h_{fs,\text{dev}} + \epsilon_{fm}^{-1} \frac{k_{f} \langle \boldsymbol{n}_{fs} \cdot \nabla \Theta_{f}  \delta_{fs} \rangle_{m}}{\langle T^{\star} \rangle_{m}^{f} - \langle T^{\star} \rangle_{m}^{s}} \exp \left( - \lambda_{T} x_{1}  \right), 
\label{eq:msheat_quasiclosure_nu}
\end{equation}
and
\begin{equation}
h_{b} \simeq h_{b,\text{dev}} \left(1- \frac{\langle \Theta \rangle_{m}^f - \langle \Theta \rangle_{m}^s }{\langle T^{\star} \rangle_{m}^{f} - \langle T^{\star} \rangle_{m}^{s}} \exp \left( - \lambda_{T} x_{1}  \right)
\right), 
\label{eq:msheat_quasiclosure_nu2}
\end{equation}
as it follows from (\ref{eq:msheat_quasitemp}). 
Here, $h_{fs,\text{dev}} \triangleq \epsilon_{fm}^{-1} \langle \boldsymbol{n}_{fs} \cdot k_{f} \nabla T^{\star}_{f}  \delta_{fs} \rangle_{m}/(\langle T^{\star} \rangle_{m}^{f} - \langle T^{\star} \rangle_{m}^{s}) $ and $h_{b,\text{dev}} \triangleq \epsilon_{fm}^{-1} \langle q_b  \delta_{s} \rangle_{m}/(\langle T^{\star} \rangle_{m}^{f} - \langle T^{\star} \rangle_{m}^{s})$ denote the macro-scale heat transfer coefficients in the developed region \cite{buckinx2016macro}, of which the last one equals $h_{\text{unit}}$ away from the channel's side walls. 
As a technical remark, we clarify that the former asymptotic expressions and definitions of $h_{fs,\text{dev}}$ and $h_{b,\text{dev}}$ have been obtained by neglecting the contributions from the spatial moments to the macro-scale temperature difference between the solid and fluid, since $  \mathrm{\nabla{T}} \cdot \left( \langle \gamma_f \boldsymbol{y} \rangle_m^f - \langle \gamma_s \boldsymbol{y} \rangle_m^s \right) \ll \langle T^{\star} \rangle_{m}^{f} - \langle T^{\star} \rangle_{m}^{s} $ for the double volume-averaging operator. 
Besides, we have $\langle \gamma_f \boldsymbol{y} \rangle_m^f = \langle \gamma_s \boldsymbol{y} \rangle_m^s =0$ in $\Omega_{\text{predev,T}} \setminus \Omega_{\text{sides}}$ anyway for the double volume-averaging operator \cite{buckinx2017macro}. 
In addition, we have neglected the contribution $ - \epsilon_{fm}^{-1} \nabla \epsilon_{fm}  \cdot  k_f\mathrm{\nabla{T}}/ (\langle T^{\star} \rangle_{m}^{f} - \langle T^{\star} \rangle_{m}^{s})$ in the definition of $h_{fs,\text{dev}}$, assuming that the porosity gradient $\nabla \epsilon_{fm}$ in the region near the side walls is perpendicular to the main flow direction $\boldsymbol{e}_s$. 

From the asymptotic expression (\ref{eq:msheat_quasiclosure_nu2}), we can determine the section $x_{1} = x_{h}$ after which the actual heat transfer coefficient $h_{b}$ agrees with the developed heat transfer coefficient $h_{\text{unit}}$ within a relative difference $\epsilon_{h}$. 
However, from our DNS experiments in Section \ref{sec:msheat_accuracy} and in particular Figures \ref{fig:msheat_Nu_x_t0.02_h0.12_s0.48_1} and \ref{fig:msheat_Nu_x_t0.02_h0.12_s0.48_2}, it appears that the exponential correction term on the right hand side of equation (\ref{eq:msheat_quasiclosure_nu2}) adds little accuracy. 
This is because it decays sufficiently fast, especially when multiplied with the temperature difference $\langle T \rangle_{m}^{f} - \langle T \rangle_{m}^{s}$ to model the macro-scale interfacial heat transfer (\ref{eq:msheat_closure_interf}). 
This is an alternative viewpoint on why we can rely on the developed correlations for $h_{\text{unit}}$ to model the macro-scale interfacial heat transfer $h_{b}$ over nearly the entire channel, except the region near the side walls.


Although our DNS experiments have indicated that the remaining closure terms for the macro-scale energy conservation equations are completely negligible within the fin array, it is still instructive to examine their form in the quasi-developed flow and heat transfer region.
The macro-scale thermal dispersion source in $\Omega_{\text{predev,T}}$ for instance is given by
\begin{equation}
\begin{aligned}
\boldsymbol{D} \simeq \boldsymbol{D}_{\text{dev}} &+ \bigl(\langle \boldsymbol{u}^{\star} \Theta \rangle_{m}  - \langle \boldsymbol{u}^{\star} \rangle_{m} \langle \Theta \rangle_{m}^{f} \bigr) 
\exp \left( - \lambda_{T} x_{1}  \right) \\
&+ \bigl( \langle \textbf{U} T^{\star} \rangle_{m} - \langle \textbf{U} \rangle_{m} \langle T^{\star} \rangle_{m}^{f} \bigr)
\exp \left( - \lambda x_{1}  \right) \\
&+ \left( \langle \textbf{U} \boldsymbol{y}\rangle_{m} - \langle \textbf{U} \rangle_m \langle \gamma_f \boldsymbol{y} \rangle_m^f \right)  \cdot \mathrm{\nabla{T}} \exp \left( - \lambda x_{1}  \right)
\,.
\end{aligned}
\label{eq:msheat_quasiclosure_D}
\end{equation}
Here, $\boldsymbol{D}_{\text{dev}} \triangleq \langle \boldsymbol{u}^{\star} T^{\star}\rangle_{m}  - \langle \boldsymbol{u}^{\star} \rangle_{m} \langle T^{\star}  \rangle_{m}^{f} + \langle \boldsymbol{u}^{\star} \boldsymbol{y}\rangle_{m} \cdot \mathrm{\nabla{T}} - \langle \boldsymbol{u}^{\star} \rangle_m \langle \gamma_f \boldsymbol{y} \rangle_m^f \cdot \mathrm{\nabla{T}} $ represents the macro-scale thermal dispersion source in $\Omega_{\text{dev,T}}$ \cite{buckinx2016macro}, while $\boldsymbol{u} = \boldsymbol{u}^{\star}  + \textbf{U} \exp \left( - \lambda x_{1}  \right)$ is the quasi-periodically developed flow field, whose mode amplitude is streamwise periodic: $\textbf{U}(\boldsymbol{x})= \textbf{U}(\boldsymbol{x} + \boldsymbol{l}_1)$ \cite{buckinx2023arxiv}. 
The first-order spatial moments of the developed velocity field $\langle \boldsymbol{u}^{\star} \boldsymbol{y}\rangle_{m}$ and the mode amplitude $\langle \textbf{U} \boldsymbol{y}\rangle_{m}$ have been notated based on the coordinate vector $ \boldsymbol{y}$ relative to the center of filter window, although this is technically an abuse of notation \cite{davit2017technical,buckinx2017macro}. 
From the asymptotic form of the thermal dispersion source (\ref{eq:msheat_quasiclosure_D}), we can argue that the contribution of its divergence is indeed negligible with respect to macro-scale heat transfer from the channel's bottom: $\rho_f c_f \nabla \cdot \boldsymbol{D} \ll \langle q_b\delta_b \rangle_m$. 
The argument relies on the length-scale (or order-of-magnitude) estimates $x_1 = O(L_1)$, $x_2 = O( l_2 )$, $\boldsymbol{u} = O(u_b)$, $T^{\star} = O(\Vert \mathrm{\nabla{T}} \Vert l_1)$, $ \boldsymbol{y}= O(l_1)$,  $\mathrm{\nabla{T}}=  O(q_b /\rho_f c_f u_b l_1)$, and the fact that $\langle u_2 \rangle_m = O( u_b l_2/L_1 )$ because of $\nabla \cdot \langle \boldsymbol{u} \rangle_m =0$. 
Besides, we have $\Vert \textbf{U} \Vert \ll  u_b$ and $\ \nabla \Theta  \ll \nabla T^{\star}$, while the spatial moments are (approximately) perpendicular to the temperature gradient $\mathrm{\nabla{T}}$. 
Based on the former estimates, we find that $\rho_f c_f \nabla \cdot \boldsymbol{D} = O \left(  \langle q_b\delta_b \rangle_m \langle u_2 \rangle_m / u_b \right) = O \left(  \langle q_b\delta_b \rangle_m l_2/ L_1 \right) $. 
Given the scale separation $l_2 \ll L_1$, and hence small lateral macro-scale velocity component $\langle u_2 \rangle_m$ \cite{vangeffelen2023developed}, the macro-scale thermal dispersion is thus negligible in $\Omega_{\text{predev,T}}$. 

Using the same length-scale estimates, it can analogously be shown that the divergence of the thermal tortuosity terms due to the solid-fluid interface and side-wall surfaces can be neglected in the quasi-developed flow and heat transfer region. 
The exact asymptotic form of these terms is given by
\begin{equation}
\begin{aligned}
\langle \boldsymbol{n}_{fs} T_{f} \delta_{fs} \rangle_{m} 
&\simeq \langle \boldsymbol{n}_{fs} T_{f}^{\star} \delta_{fs} \rangle_{m} +
\langle  \boldsymbol{n}_{fs} \boldsymbol{y} \delta_{fs} \rangle_{m} \cdot \mathrm{\nabla{T}}  \\
& \qquad + \langle  \boldsymbol{n}_{fs} \delta_{fs} \rangle_{m}  \mathrm{\nabla{T}} \cdot \boldsymbol{x} +
\langle \boldsymbol{n}_{fs} \Theta_{f} \delta_{fs} \rangle_{m} \exp \left( - \lambda_{T} x_{1}  \right) \,, \\
\langle \boldsymbol{n} T_{f} \delta_{\text{sides},f} \rangle_{m} 
&\simeq \langle \boldsymbol{n} T_{f}^{\star} \delta_{\text{sides},f} \rangle_{m} +
\langle  \boldsymbol{n}_{fs} \boldsymbol{y} \delta_{\text{sides},f} \rangle_{m} \cdot \mathrm{\nabla{T}}  \\
&\qquad + \langle  \boldsymbol{n} \delta_{\text{sides},f} \rangle_{m}  \mathrm{\nabla{T}} \cdot \boldsymbol{x} +
\langle \boldsymbol{n} \Theta_{f} \delta_{\text{sides},f}\rangle_{m} \exp \left( - \lambda_{T} x_{1}  \right) \, 
\end{aligned}
\label{eq:msheat_quasiclosure_tort}
\end{equation}
where $\langle  \boldsymbol{n}_{fs} \delta_{fs} \rangle_{m} = - \nabla \epsilon_{fm}$ and 
$\langle  \boldsymbol{n}_{fs}  \boldsymbol{y} \delta_{fs} \rangle_{m} = - \nabla \langle  \boldsymbol{y} \gamma_{f} \rangle_{m}$, as shown in \cite{buckinx2017macro}.

We add that the macro-scale dispersion and tortuosity terms in the quasi-developed region can also be represented by means of an exact dispersion tensor and tortuosity tensor, as we show in \ref{sec:temperaturemap}.
As such, these terms can be modelled via a gradient-diffusion law, as often hypothesized in the macro-scale models for porous media, see for instance \cite{quintard1997two}.


We remark that in principle, the weighting function $m$ of the filter operator $\langle \; \rangle_m$ should be matched to both eigenvalues $\lambda_{T}$ and $\lambda$, as discussed in \cite{buckinx2015macro, buckinx2017macro}. 
Otherwise, the exponential terms in (\ref{eq:msheat_quasiclosure_nu})-(\ref{eq:msheat_quasiclosure_tort}) cannot be moved outside the filter operator $\langle \; \rangle_m$. 
Nevertheless, for most cases, the double volume-averaging operator (\ref{eq:weightingfunction}) is accurate enough to ensure the validity of the asymptotic forms (\ref{eq:msheat_quasiclosure_nu})-(\ref{eq:msheat_quasiclosure_tort}), especially because we have $\lambda_{T} l_{1} \ll 1$ when the Péclet number is larger than 100 according to Figures \ref{fig:msheat_lambdaT_PE_t}-\ref{fig:msheat_lambdaT_s}. 



\begin{figure}[ht!]
\centering
\includegraphics[scale = 0.75]{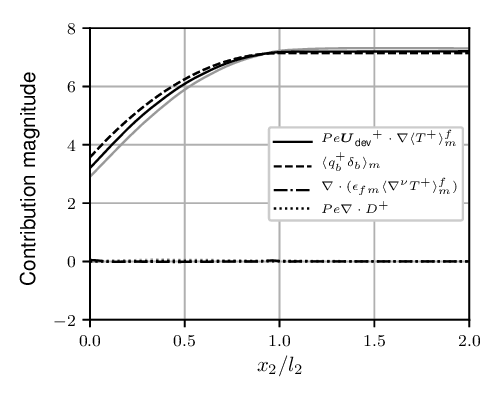}
\caption{\label{fig:closure_y} Macro-scale energy balance along the channel width at $x_{1} = 18 l_{1}$ (black) and $x_{1} = 2 l_{1}$ (grey), when $Re_{b}=28$,  $Pr_{f}=7$, $h/l=0.12$, $s/l=0.48$, $t/l=0.02$, $s_{0}=l_{1}$, $s_{N}=2.5 l_{1}$, $N_{1}=20$, and $N_{2}=10$}
\end{figure}

Due to the high Péclet number of the fluid ($Pe \gg 1$) in most applications, we can typically neglect the thermal diffusion in the fluid.
Furthermore, due to the high conductivity of the solid ($k_s/k_f \gg 100$), also the temperature gradients in the solid are typically very small.
Consequently, the macro-scale energy conservation equations in $\Omega_{\text{predev,T}} \cup \Omega_{\text{dev,T}}$ can be simplified into 
\begin{equation}
\begin{aligned}
  \rho_{f} c_{f} \boldsymbol{U}_{\text{dev}} \cdot \nabla \langle T \rangle_{m}^{f} &\simeq   - \langle q_{b} \delta_{bf} \rangle_{m} + k_{f} \langle \boldsymbol{n}_{fs} \cdot \nabla T_{f} \delta_{fs} \rangle_{m}, \\
  0 &\simeq - \langle q_{b} \delta_{bs} \rangle_{m} - k_{s} \langle \boldsymbol{n}_{fs} \cdot \nabla T_{s} \delta_{fs} \rangle_{m}.
\end{aligned}
\label{eq:msheat_TemperatureMSside}
\end{equation}
Conform to the preceding literature \cite{buckinx2022arxiv}, $\boldsymbol{U}_{\text{dev}} \triangleq \langle \boldsymbol{u}^{\star} \rangle_{m} \triangleq  U_{\text{dev}}(x_2) \boldsymbol{e}_{1}$ denotes here the developed macro-scale velocity profile, which is uniform away from the side walls of the channel, as characterized in \cite{vangeffelen2023developed}. 
As a result, the overall macro-scale energy balance in the region $\Omega_{\text{predev,T}} \cup \Omega_{\text{dev,T}}$ becomes
\begin{equation}
\begin{aligned}
\rho_{f} c_{f} \boldsymbol{U}_{\text{dev}} \cdot \nabla \langle T \rangle_{m}^{f} & \simeq
- \langle q_{b} \delta_{b} \rangle_{m}  = -h_{b} \epsilon_{fm} \left( \langle T \rangle_{m}^{f} - \langle T \rangle_{m}^{s} \right),
\end{aligned}
\label{eq: final macro-scale temperature equation quasi-developed and developed region}
\end{equation}
with $h_{b} \simeq h_{b,\text{dev}}$. 
Figure \ref{fig:closure_y} illustrates the validity of this final macro-scale energy balance for a section in the developed region ($x_{1} = 18 l_{1}$) as well as one in the quasi-developed region ($x_{1} = 2 l_{1}$). 
In this figure, the non-dimensional macro-scale velocity, uniform heat flux and macro-scale dispersion are defined such that $\boldsymbol{U}_{\text{dev}}^{+} \triangleq \boldsymbol{U}_{\text{dev}}/u_{b}$, $q_{b}^{+} \triangleq q_{b}/q_{b} = 1$ and $\boldsymbol{D}^{+} \triangleq \boldsymbol{D} (q_b l_1)/(k_f u_{b})$, while $\nabla$ and $\delta_{b}$ denote the dimensionless gradient operator and Dirac surface operator based on the reference length $l_1$. 
The main takeaway is that according to the final macro-scale energy balance (\ref{eq: final macro-scale temperature equation quasi-developed and developed region}), the developed heat transfer coefficient $h_{b,\text{dev}}$ is the only relevant closure coefficient required to reconstruct the macro-scale temperature profiles of the fluid and solid over almost the entire channel. 
Although the former heat transfer coefficient is completely known in the core of the channel, where it becomes uniform (i.e. $h_{b,\text{dev}}= h_{b,\text{unit}}$), its spatial behavior closer to the side walls has not yet been investigated.
This will be the topic of the next section. 

\newpage
\clearpage

\section{\label{sec:msheat_sidewall}Influence of the side-wall region on macro-scale heat transfer}

Within a certain distance $l_{\text{sides,T}}$ from the channel side walls in the developed heat transfer region $\Omega_{\text{dev,T}}$, the streamwise periodic temperature field $T^{\star}$ will be no longer periodic along the transversal direction $\boldsymbol{e}_{2}$. 
This is a consequence of the no-heat-flux condition at the channel boundary $\Gamma_{\text{sides}}$, and the fact that the velocity field loses its transversal periodicity close to side walls, due to the no-slip condition at $\Gamma_{\text{sides}}$ \cite{vangeffelen2023developed}. 
For our final macro-scale model (\ref{eq: final macro-scale temperature equation quasi-developed and developed region}), this implies that macro-scale heat transfer coefficient $h_{b}$ in $\Omega_{\text{dev,T}}$
will vary with the coordinate $x_2$ near the side walls, since $\langle T^{\star} \rangle_{m}^{f}$ and $\langle T^{\star} \rangle_{m}^{s}$ are no longer spatially constant in that region. 
In order to specify the developed macro-scale heat transfer coefficient $h_{b,\text{dev}}$ in $\Omega_{\text{dev,T}}$, the profile of the temperature difference $\langle T \rangle_{m}^{f}-\langle T \rangle_{m}^{s} \simeq  \langle T^{\star} \rangle_{m}^{f}-\langle T^{\star} \rangle_{m}^{s} $ within a distance $l_{\text{sides,T}}$ from the side walls in $\Omega_{\text{dev,T}}$ thus needs to be known. 



According to our DNS data, the distance $l_{\text{sides,T}}$ is smaller than the lateral size $l_{2}$ of the unit cell, when the width of the channel is sufficiently large to allow for transversal periodicity of the velocity field. 
Consequently, the side-wall region where the heat transfer coefficient $h_{b,\text{dev}}$ varies with $x_2$ practically corresponds to $\Omega_{\text{sides,T}} = \{ \boldsymbol{x} \in \Omega | x_{2} \in (0, l_{2}) \cup (L_{2} - l_{2}, L_{2} ) \}$, which is the region where 
the lateral porosity gradient $d\epsilon_{fm}/dx_2$ is nonzero. 
Conversely, the region where the heat transfer coefficient and other macro-scale closure terms are uniform, is given by $\Omega_{\text{uniform,T}} = \{ \boldsymbol{x} \in \Omega | x_{1} \in (x_{\text{dev,T}},x_{\text{out}}), x_{2} \in (l_{2},L_{2} - l_{2} ) \}$. 

\begin{figure}[ht!]
\centering
\begin{minipage}{.475\textwidth}
\hspace{-5mm}
\includegraphics[scale=0.75]{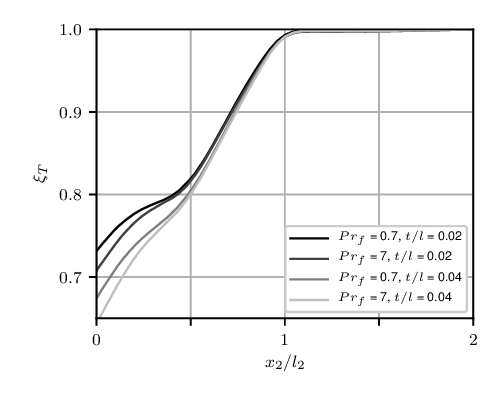}
\caption{\label{fig:xi_T_y} Macro-scale temperature difference profile in the side-wall region, when $Re_{b} l / (2 L_{3}) = \rho_{f} u_{b} l / \mu_{f} = 100$, $N_{2} = 10$, $h/l=0.12$, $s/l=0.48$ \newline\newline}
\end{minipage}
\hfill
\begin{minipage}{.475\textwidth}
\hspace{-5mm}
\includegraphics[scale = 0.75]{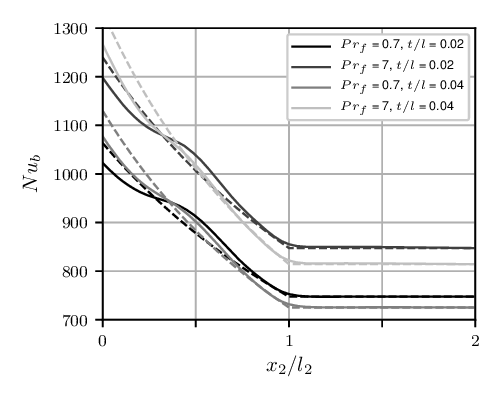}
\caption{\label{fig:Nu_y} Macro-scale heat transfer coefficient $Nu_{b,\text{dev}}$ (full) and its prediction $\xi_{T}^{-1} Nu_{\text{unit}}$ (dashed) by the developed correlation from \cite{vangeffelen2022nusselt}, when $Re_{b} l / (2 L_{3}) = \rho_{f} u_{b} l / \mu_{f} = 100$, $N_{2} = 10$, $h/l=0.12$, $s/l=0.48$ \newline}
\end{minipage}
\end{figure}

To model the macro-scale heat transfer coefficient $h_{b,\text{dev}}$ in $\Omega_{\text{sides,T}} \cap \Omega_{\text{dev,T}}$, we represent the profile of the macro-scale temperature difference in this region by the following shape function $\xi_{T}$: 
\begin{equation}
\xi_{T} ( x_{2} ) \triangleq \frac{\left( \langle T \rangle_{m}^{f}-\langle T \rangle_{m}^{s} \right) \Big\rvert_{x_{2}}} {\quad \left( \langle T \rangle_{m}^{f}-\langle T \rangle_{m}^{s} \right) \Big\rvert_{x_{2}=l_{2}}}\,.
\end{equation}
The non-dimensional parameter $\xi_{T}$ is essentially a closure variable that maps the local macro-scale temperature difference between the fluid and solid to its uniform value outside the side-wall region. 
The variation of $\xi_{T}$ with the coordinate $x_2$ perpendicular to each side wall is illustrated in Figure \ref{fig:xi_T_y}. 
For the data considered in Figure \ref{fig:xi_T_y}, we find that $\xi_{T}$ can be approximated by a linear profile within a maximum relative error of 5\%: 
\begin{equation}
\xi_T \left( x_{2} \right) \simeq
\begin{cases}
1 - \kappa (x_{2} - l_{2})          & \text{for } x_{2} \in (0, l_{2}), \\
1                                   & \text{for } x_{2} \in (l_{2}, L_{2}-l_{2}), \\
1 - \kappa' (L_{2} - x_{2} - l_{2}) & \text{for } x_{2} \in (L_{2}-l_{2}, L_{2}).  
\end{cases}
\label{eq: linear profile macro-scale temperature difference in side-wall}
\end{equation}
The linear shape function $\xi_{T}$ is then determined by the constant slopes $\kappa$ and $\kappa'$ at each side wall of the channel, which depends on the Reynolds number, the material properties, and fin-row geometry. 
Although the slopes $\kappa$ and $\kappa'$ of the linear profile do not necessarily have to be equal because of the asymmetric shape of the offset strip fins, we did find that the assumption $\kappa \simeq \kappa'$ is justified for all of the investigated cases. 
For example, the following correlations capture the data of $\kappa$ and $\kappa'$ within a relative error of 1\% for two channel geometries with $N_{2} = 10$, $h/l=0.12$, $s/l=0.48$, $Pr_{f}=0.7$ and $Re_{b} \in (28,192)$: $\kappa \simeq 2.23e\text{-}4 Re_{b} + 0.291$ when $t/l=0.02$ and $\kappa \simeq 3.64e\text{-}4 Re_{b} + 0.351$ when $t/l=0.04$. 

By definition, the shape function $\xi_{T}$ can be employed to model the local value of the macro-scale heat transfer coefficient $h_{b}$ from its developed prediction $h_{\text{unit}}$: 
\begin{equation}
    h_{b\text{,dev}} = \xi_{T}^{-1} h_{\text{unit}} \,.
    \label{eq:htcsidewall}
\end{equation}
The macro-scale heat transfer coefficient $h_{\text{unit}}$ can be computed from the developed Nusselt number correlation (\ref{eq: nusselt number developed correlations}) based on the uniform developed macro-scale velocity outside the side-wall region, i.e.  $\langle \boldsymbol{u}^{\star} \rangle_m = Ue_1$ at $x_2=l_2$. 
Just like the macro-scale velocity profile $\xi$ in the side-wall region, also the temperature profile $\xi_T$ can be determined by resolving the flow and temperature field on an extended unit cell, as suggested in \cite{vangeffelen2023developed}. 
Such an extended unit cell consists of one unit cell in $\Omega_{\text{sides,T}}$ and another adjacent unit cell in $\Omega_{\text{uniform,T}}$, as the width of the side-wall region equals $l_2$. 

In Figure \ref{fig:Nu_y}, we have illustrated the accuracy of our linear model (\ref{eq: linear profile macro-scale temperature difference in side-wall}) for the developed heat transfer coefficient (\ref{eq:htcsidewall}) in the side-wall region. 
Hereto, we compared the predicted value for $h_{b, \text{dev}}$ according to our linear model (\ref{eq:htcsidewall}) with its actual value, which we computed from definition (\ref{eq:msheat_Nub}). 
Both the predicted and actual values have been represented in the form of a dimensionless Nusselt number: $Nu_{b,\text{dev}} \triangleq h_{b,\text{dev}} l^{2}/k_{f}$. 
According to Figure \ref{fig:Nu_y}, the inexact linear expressions for $\xi_T$  result in a relative mean and maximum error of 5\% and 11\% in $\Omega_{\text{sides,T}} \cap \Omega_{\text{dev,T}}$.


Finally, we add that the macro-scale temperature profile $\xi_{T}$ also allows us to determine the effective boundary conditions for the macro-scale fluid temperature governed by equations (\ref{eq:msheat_TemperatureMSside}). 
At least, this is true when the conductivity of the solid material is so high that the temperature in the solid region $\Omega_s$ is virtually constant, as we then have $\boldsymbol{n} \cdot \nabla \langle T \rangle_{m}^{f} \gg \boldsymbol{n} \cdot \nabla \langle T \rangle_{m}^{s}$, and thus
\begin{equation}
\begin{aligned}
  \boldsymbol{n} \cdot \nabla \langle T \rangle_{m}^{f} &\simeq \boldsymbol{n} \cdot \nabla \xi_{T} \left( \langle T \rangle_{m}^{f}-\langle T \rangle_{m}^{s} \right) \Big\rvert_{x_{2}=l_{2}} = -\kappa \left( \langle T \rangle_{m}^{f}-\langle T \rangle_{m}^{s} \right) \Big\rvert_{x_{2}=l_{2}}, \\
  \boldsymbol{n} \cdot \nabla \langle T \rangle_{m}^{s} &\simeq 0, 
\end{aligned}
\label{eq:msheat_BC_simp}
\end{equation}
for $\boldsymbol{x} \in \Gamma_{\text{sides}} \cap \Omega_{\text{dev,T}}$. 
Consequently, the constant $(k_{f} \kappa)$ can be interpreted as an effective heat transfer coefficient relating the effective macro-scale heat flux at the side wall of the channel to the uniform macro-scale temperature difference just outside the side-wall region. 


\newpage
\clearpage

\section{\label{sec:msheat_conclusion}Conclusion}

In this work, we have presented a macro-scale description of the developed and quasi-developed heat transfer regimes in micro- and mini-channels with arrays of periodic offset strip fins, subject to a uniform heat flux. 
The theoretical framework has been supported by direct numerical simulations of the developing temperature and flow fields in typical channel geometries. 
Both air and water flows have been considered for Reynolds numbers ranging from 28 to 1224. 
The macro-scale temperature fields have been calculated by repeated volume-averaging of the obtained temperature fields over each local unit cell in the channel.

We observed that the onset point of the developed macro-scale heat transfer regime scales linearly with the Péclet number for a given offset strip fin channel geometry. 
Yet, for the cases considered in this work, the thermal development length remains limited with respect to the total length of the array used in typical micro- and mini-channel applications. 
Particularly for air, the macro-scale temperature field can be considered developed after two unit cells in the offset strip fin array. 
Consequently, the developed Nusselt number correlations from our previous work are able to accurately model the macro-scale heat transfer coefficient throughout almost the entire channel.
They predict the macro-scale temperature difference between the fluid and solid material near the channel inlet within a mean relative error of 10\%.

Similar to the macro-scale flow field, the macro-scale temperature field 
becomes quasi-developed virtually after the first unit cell in the fin array for all the cases in this work.
So, the macro-scale temperature field in the region immediately upstream of the developed region is characterized by a single dominant mode, which decays exponentially along the main flow direction. 
The relatively large eigenvalues of this mode explain the quick onset of the developed heat transfer regime, as well as the scaling of its onset point with respect to the Reynolds number, Péclet number and aspect ratio.
Moreover, the large eigenvalues are directly responsible for the good accuracy of the developed heat transfer (or Nusselt number) correlations in the developing region.
The small mode amplitudes still account for some of the secondary effects of the unit cell geometry on the development length. 


Finally, we have empirically shown that the influence of the channel's side walls on the macro-scale heat transfer coefficient can be captured by a shape function, which varies linearly with the distance from the side walls to a first approximation.
Through this shape function, the local macro-scale heat transfer coefficient in the side-wall region can be modeled with a mean and maximum error of 5\% and 11\%, according to our data. 

\section{\label{sec:contributions}Contributions}
The macro-scale description based on the double volume-averaging operation and the validation of its computational framework were carried out by G. Buckinx. 
G. Buckinx formulated the theoretical framework and modelling equations for the quasi-developed macro-scale heat transfer regime. 
A. Vangeffelen carried out all numerical heat transfer simulations and post-processing calculations. 
The results were interpreted by A. Vangeffelen, with input from G. Buckinx regarding the existing literature. 
The paper was written by A. Vangeffelen and G. Buckinx, with input from C. De Servi, M. R. Vetrano, and M. Baelmans. 

\section{\label{sec:acknowledgements}Acknowledgements}
The work documented in this paper was funded by the Flemish Institute for Technological Research (VITO) through the Ph.D. grant 1810603 of A. Vangeffelen, and by the Research Foundation — Flanders (FWO) through the post-doctoral project grant 12Y2919N of G. Buckinx. 
The VSC (Flemish Supercomputer Center), funded by the Research Foundation - Flanders (FWO) and the Flemish Government, supplied the resources and services used in this work.


\newpage
\clearpage

\appendix

\section{\label{sec:temperaturemap} Closure problem for the quasi-periodically developed heat transfer regime under a uniform heat flux}

\subsection{Closure for the periodically developed region}
If we propose the closure mapping
\begin{equation}
\begin{aligned}
T_{f}^{\star} = \boldsymbol{\psi}_{f} \cdot \mathrm{\nabla{T}}\,, &&
T_{s}^{\star} = \boldsymbol{\psi}_{s} \cdot \mathrm{\nabla{T}} \,,
\end{aligned}
\label{eq: developed region closure mapping}
\end{equation}
and substitute this mapping in the periodic temperature equations for the periodically developed heat transfer regime under a uniform heat flux \cite{vangeffelen2022nusselt}, we obtain the following problem for the closure variable $\boldsymbol{\psi}$ in the developed heat transfer region:
\begin{equation}
\begin{aligned}
  \rho_{f} c_{f} \boldsymbol{u}^{\star}_f \cdot \nabla \boldsymbol{\psi}_{f} + \rho_{f} c_{f} \boldsymbol{u}^{\star}_f  &=  k_{f} \nabla^2 \boldsymbol{\psi}_{f} && \text{in } \Omega_{f}, \\
  0 &= k_{s} \nabla^2 \boldsymbol{\psi}_{s} && \text{in } \Omega_{s}, \\
  \boldsymbol{\psi} \left( \textbf{x} + \textbf{l}_{1}  \right) &= \boldsymbol{\psi} \left( \textbf{x}  \right)\,, & & \\
  - \boldsymbol{n} \cdot k_{f} \left( \nabla \boldsymbol{\psi}_{f} + \boldsymbol{I} \right) &= - \rho_{f} c_{f} \boldsymbol{U}_{\text{dev}} / \langle \delta_{b} \rangle_{m} & &\text{in } \Gamma_{bf}, \\
  - \boldsymbol{n} \cdot k_{s} \left( \nabla \boldsymbol{\psi}_{s} + \boldsymbol{I} \right) &= - \rho_{f} c_{f} \boldsymbol{U}_{\text{dev}} / \langle \delta_{b} \rangle_{m} & &\text{in } \Gamma_{bs},  \\
  - \boldsymbol{n} \cdot \left( \nabla \boldsymbol{\psi} + \boldsymbol{I} \right) &= 0 & &\text{in } \Gamma_{t} \cup \Gamma_{\text{sides}}, \\
  \boldsymbol{\psi}_{f} &= \boldsymbol{\psi}_{s} & &\text{in } \Gamma_{fs}, \\
  - \boldsymbol{n}_{fs} \cdot k_{f} \left( \nabla \boldsymbol{\psi}_{f} + \boldsymbol{I} \right) &= - \boldsymbol{n}_{fs} \cdot k_{s} \left( \nabla \boldsymbol{\psi}_{s} + \boldsymbol{I} \right) & &\text{in } \Gamma_{fs}. \\
\end{aligned}
\label{eq: developed region closure problem}
\end{equation}

Through the closure mapping (\ref{eq: developed region closure mapping}), the gradients of the macro-scale temperatures in the fluid and solid given by (\ref{eq:msheat_MSTdev}) can be mapped to the constant temperature gradient in the developed regime: $\nabla \langle T \rangle_{m}^{f} = \boldsymbol{B}_{fm} \cdot \mathrm{\nabla{T}}$ and $\nabla \langle T \rangle_{m}^{s} = \boldsymbol{B}_{sm} \cdot \mathrm{\nabla{T}}$. 
The mapping tensors are defined as $\boldsymbol{B}_{fm} \triangleq \boldsymbol{I} + \nabla \langle \gamma_f \boldsymbol{y} \rangle_{m}^{f} + \nabla \langle \boldsymbol{\psi} \rangle_{m}^{f}$ and $\boldsymbol{B}_{sm} \triangleq \boldsymbol{I} + \nabla \langle \gamma_s \boldsymbol{y} \rangle_{m}^{s} + \nabla \langle \boldsymbol{\psi} \rangle_{m}^{s}$. 
Outside the side-wall region, both $\boldsymbol{B}_{fm}$ and $\boldsymbol{B}_{sm}$ become equal to the identity tensor $\boldsymbol{I}$. 

Consequently, the closure problem (\ref{eq: developed region closure problem}) allows us to express all closure terms for the macro-scale temperature equations in terms of the macro-scale temperature gradients $\nabla \langle T \rangle_{m}^{f}$ and $\nabla \langle T \rangle_{m}^{s}$. 
For the macro-scale interfacial heat transfer, we obtain for instance
\begin{equation}
\begin{aligned}
     \langle q_{fs} \delta_{fs} \rangle_{m} &= - \boldsymbol{K}^{f}_{fs,\text{dev}}  \cdot \nabla \langle T \rangle_{m}^{f} = - \boldsymbol{K}^{s}_{fs,\text{dev}} \cdot \nabla \langle T \rangle_{m}^{s}\, 
\end{aligned}
\end{equation}
with 
\begin{equation}
\begin{aligned}
\boldsymbol{K}^{f}_{fs,\text{dev}} &\triangleq k_{f} \langle \boldsymbol{n}_{fs} \cdot \left( \nabla \boldsymbol{\psi}_{f} + \boldsymbol{I} \right) \delta_{fs} \rangle_{m} \cdot \boldsymbol{B}_{fm}^{-1} \,, \\
\boldsymbol{K}^{s}_{fs,\text{dev}} & \triangleq 
k_{s} \langle \boldsymbol{n}_{fs} \cdot \left( \nabla \boldsymbol{\psi}_{s} + \boldsymbol{I} \right) \delta_{fs} \rangle_{m} \cdot \boldsymbol{B}_{sm}^{-1}\,.
\end{aligned}
\end{equation}
Likewise, the thermal dispersion source can be written as
\begin{equation}
    \boldsymbol{D} =  \boldsymbol{K}_{d,\text{dev}} \cdot \nabla \langle T \rangle_{m}^{f} \,, 
\end{equation}
with the developed dispersion tensor defined as 
\begin{equation}
    \boldsymbol{K}_{d,\text{dev}} \triangleq \left( \langle \boldsymbol{u}^{\star} \boldsymbol{\psi}_{f} \rangle_{m} - \epsilon_{fm} \langle \boldsymbol{u}^{\star} \rangle_{m}^{f} \langle \boldsymbol{\psi}_{f} \rangle_{m}^{f} + \langle \boldsymbol{u}^{\star} \boldsymbol{y}\rangle_{m} - \langle \boldsymbol{u}^{\star} \rangle_{m} \langle \gamma_f \boldsymbol{y} \rangle_{m}^{f} \right) \cdot \boldsymbol{B}_{fm}^{-1} \,. 
\end{equation}
Lastly, the thermal tortuosity can be represented as
\begin{equation}
    \langle \boldsymbol{n}_{fs} T_{f} \delta_{fs} \rangle_{m} = \boldsymbol{K}_{t,\text{dev}} \cdot \nabla \langle T \rangle_{m}^{f} \,, 
\end{equation}
with the developed tortuosity tensor defined as
\begin{equation}
\boldsymbol{K}_{t,\text{dev}} \triangleq ( \langle \boldsymbol{n}_{fs} \boldsymbol{\psi}_{f} \delta_{fs} \rangle_{m} + \langle \boldsymbol{n}_{fs} \boldsymbol{y} \delta_{fs} \rangle_{m} + \langle  \boldsymbol{n}_{fs} \delta_{fs} \rangle_{m} \boldsymbol{x} ) \cdot \boldsymbol{B}_{fm}^{-1} \,.
\end{equation}
A similar representation of the tortuosity terms due to the channel's side walls may be readily deduced.

\subsection{Closure for the quasi-periodically developed region}
If we propose the closure mapping
\begin{equation}
\begin{aligned}
\Theta_{f}= \boldsymbol{\Psi}_{f} \cdot \mathrm{\nabla{T}} \,, &&
\Theta_{s} = \boldsymbol{\Psi}_{s} \cdot \mathrm{\nabla{T}} \,,
\end{aligned}
\label{eq: quasideveloped region closure mapping}
\end{equation}
 and substitute this mapping in the eigenvalue problem for the quasi-periodically developed heat transfer regime under a uniform heat flux \cite{buckinx2024arxiv}, we obtain the following problem for the closure variable $\boldsymbol{\Psi}$ in the quasi-developed region:
\begin{equation}
\begin{aligned}
  \rho_{f} c_{f} \boldsymbol{u}^{\star}_f \cdot \left(\nabla \boldsymbol{\Psi}_{f}  - \boldsymbol{\lambda}_{T} \boldsymbol{\Psi}_{f} \right) &= k_{f} \nabla^2 \boldsymbol{\Psi}_{f} - \left( 2 k_{f} \nabla \boldsymbol{\Psi}_{f} \right) \cdot \boldsymbol{\lambda}_{T} + k_{f} \boldsymbol{\Psi}_{f} \lambda_{T}^{2} &&\quad \text{in } \Omega_{f} \,,\\
   0 &= k_{s} \nabla^2 \boldsymbol{\Psi}_{s} - \left( 2 k_{s} \nabla \boldsymbol{\Psi}_{s} \right) \cdot \boldsymbol{\lambda}_{T}+ k_{s} \boldsymbol{\Psi}_{s} \lambda_{T}^{2} &&\quad \text{in } \Omega_{s} \,, \\  
  \boldsymbol{\Psi} \left( \boldsymbol{x} + \boldsymbol{l}_{1}  \right) &= \boldsymbol{\Psi} \left( \boldsymbol{x}  \right) \,, && \\
  - \boldsymbol{n} \cdot \left( \nabla \boldsymbol{\Psi}  - \boldsymbol{\lambda}_{T} \boldsymbol{\Psi}  \right) &= 0 && \quad \text{in } \Gamma_{b} \cup \Gamma_{t} \cup \Gamma_{\text{sides}} \,, \\
  \boldsymbol{\Psi}_{f} &= \boldsymbol{\Psi}_{s} && \quad \text{in } \Gamma_{fs}, \\
  - \boldsymbol{n}_{fs} \cdot k_{f} \left( \nabla \boldsymbol{\Psi}_{f} -   \boldsymbol{\lambda}_{T}  \boldsymbol{\Psi}_{f}  \right) &= - \boldsymbol{n}_{fs} \cdot k_{s} \left( \nabla \boldsymbol{\Psi}_{s} -  \boldsymbol{\lambda}_{T} \boldsymbol{\Psi}_{s} \right) && \quad \text{in } \Gamma_{fs} \,.
\end{aligned}
\label{eq: closure problem quasi-developed region}
\end{equation}
In the former closure problem, the eigenvector $\boldsymbol{\lambda}_{T} \triangleq \lambda_{T} \boldsymbol{e}_s$ has a different sign than in the original eigenvalue problem from \cite{buckinx2024arxiv}.
To find a unique solution for the closure variable $\boldsymbol{\Psi}$, its averaged value over a transversal row of the array may be imposed:
\begin{equation}
    \langle \boldsymbol{\Psi} \rangle_{\text{row}} \cdot \boldsymbol{e}_s = \Psi_{0}\,.  
\end{equation}
Here, the row-averaging operator $\langle \; \rangle_{\text{row}} $ is defined as in \cite{buckinx2022arxiv}, and the scalar $\Psi_{0}$ determines the temperature perturbation size $\epsilon_{0,T}$. 

The solution to the closure problem (\ref{eq: closure problem quasi-developed region}) enables us again to express all closure terms in terms of the macro-scale temperature gradients $\nabla \langle T \rangle_{m}^{f}$ and $\nabla \langle T \rangle_{m}^{s}$.
The macro-scale interfacial heat transfer in the quasi-developed region is for instance given by
\begin{equation}
\begin{aligned}
 \langle q_{fs} \delta_{fs} \rangle_{m} &= - \boldsymbol{K}^{f}_{fs,\text{quasi-dev}}  \cdot \nabla \langle T \rangle_{m}^{f} = - \boldsymbol{K}^{s}_{fs,\text{quasi-dev}} \cdot \nabla \langle T \rangle_{m}^{s}\,
\end{aligned}
\end{equation}
with 
\begin{equation}
\begin{aligned}
\boldsymbol{K}^{f}_{fs,\text{quasi-dev}} &\triangleq \boldsymbol{K}^{f}_{fs,\text{dev}} +
k_{f} \langle \boldsymbol{n}_{fs} \cdot \left( \nabla \boldsymbol{\Psi}_{f} - \boldsymbol{\lambda}_{T} \boldsymbol{\Psi}_{f}  \right) \delta_{fs} \rangle_{m} \cdot \boldsymbol{B}_{fm}^{-1}  \exp \left( - \boldsymbol{\lambda}_{T}  \cdot \boldsymbol{x} \right) \,, \\
\boldsymbol{K}^{s}_{fs,\text{quasi-dev}} & \triangleq \boldsymbol{K}^{s}_{fs,\text{dev}} + k_{s} \langle \boldsymbol{n}_{fs} \cdot \left( \nabla \boldsymbol{\Psi}_{s} - \boldsymbol{\lambda}_{T} \boldsymbol{\Psi}_{s}  \right) \delta_{fs} \rangle_{m} \cdot \boldsymbol{B}_{sm}^{-1}  \exp \left( - \boldsymbol{\lambda}_{T}  \cdot \boldsymbol{x} \right) \,.
\end{aligned}
\end{equation}
Further, the thermal dispersion source is given by
\begin{equation}
\boldsymbol{D} = \boldsymbol{K}_{d,\text{quasi-dev}} \cdot \nabla \langle T \rangle_{m}^{f} \,,
\end{equation}
where the thermal dispersion tensor in the quasi-developed region is defined as
\begin{equation}
\begin{aligned}
    \boldsymbol{K}_{d,\text{quasi-dev}} \triangleq 
\boldsymbol{K}_{d,\text{dev}}
&+ \bigl[ \left( \langle \boldsymbol{u}^{\star} \boldsymbol{\Psi}_{f} \rangle_{m}  - \langle \boldsymbol{u}^{\star} \rangle_{m} \langle \boldsymbol{\Psi}_{f} \rangle_{m}^{f} \right) \exp \left( - \boldsymbol{\lambda}_{T}  \cdot \boldsymbol{x} \right)\\
    & \qquad + ( \langle \textbf{U} \boldsymbol{\psi}_{f} \rangle_{m} - \langle \textbf{U} \rangle_{m} \langle \boldsymbol{\psi}_{f} \rangle_{m}^{f} \\
    & \qquad + \langle \textbf{U} \boldsymbol{y}\rangle_{m} - \langle \textbf{U} \rangle_m \langle \gamma_f \boldsymbol{y} \rangle_m^f ) \exp \left( - \boldsymbol{\lambda}_{T}  \cdot \boldsymbol{x} \right) \bigr] \cdot \boldsymbol{B}_{fm}^{-1} \,. \\
\end{aligned}
\end{equation}
We remark that the velocity fields $\textbf{U}$ and $\boldsymbol{u}^{\star}$ can be further related to the (uniform) macro-scale velocity $\boldsymbol{U}_{\text{dev}} \triangleq \langle \boldsymbol{u}^{\star}\rangle_m$ in the quasi-developed flow region via the closure problems presented in \cite{buckinx2022arxiv}.

Analogously, we can express the thermal tortuosity due to the fluid-solid interface as
\begin{equation}
    \langle \boldsymbol{n}_{fs} T_{f} \delta_{fs} \rangle_{m} = \boldsymbol{K}_{t,\text{quasi-dev}} \cdot \nabla \langle T \rangle_{m}^{f} \,,
\end{equation}
by defining the thermal tortuosity tensor in the quasi-developed region as
\begin{equation}
\boldsymbol{K}_{t,\text{quasi-dev}} \triangleq \boldsymbol{K}_{t,\text{dev}} + \langle \boldsymbol{n}_{fs} \boldsymbol{\Psi}_{f} \delta_{fs} \rangle_{m}  \cdot \boldsymbol{B}_{fm}^{-1}  \exp \left( - \boldsymbol{\lambda}_{T}  \cdot \boldsymbol{x} \right) \,.
\end{equation}

Finally, we can also obtain a relationship between the deviation parts of the fluid and solid temperature on one hand, and the macro-scale temperature gradients on the other hand: 
\begin{equation}
\begin{aligned}
    \widetilde{T_{f}} &= \left( \widetilde{\boldsymbol{\psi}}_{f} - \langle \gamma_f \boldsymbol{y} \rangle_{m}^{f} + \widetilde{\boldsymbol{\Psi}}_{f} \exp \left( - \boldsymbol{\lambda}_{T}  \cdot \boldsymbol{x} \right) \right) \cdot \boldsymbol{B}_{fm}^{-1} \cdot \nabla \langle T \rangle_{m}^{f} \,, \\
    \widetilde{T_{s}} &= \left( \widetilde{\boldsymbol{\psi}}_{s} - \langle \gamma_f \boldsymbol{y} \rangle_{m}^{s} + \widetilde{\boldsymbol{\Psi}}_{s} \exp \left( - \boldsymbol{\lambda}_{T}  \cdot \boldsymbol{x} \right) \right) \cdot \boldsymbol{B}_{sm}^{-1} \cdot \nabla \langle T \rangle_{m}^{s} \, . 
\end{aligned}
\label{eq: closure deviation fields quasi-developed region}
\end{equation}
Here, the deviation operator is defined as $\widetilde{\boldsymbol{\phi}} \triangleq \boldsymbol{\phi}  - \langle \boldsymbol{\phi} \rangle^{f}_m\gamma_f - \langle \boldsymbol{\phi} \rangle^{s}_m \gamma_s $.
The former temperature deviation parts (\ref{eq: closure deviation fields quasi-developed region}) are not of interest for this work because we rely on the exact definitions of the closure terms.
Nevertheless, they play an important role in the available closure problems for porous media, see for instance \cite{quintard1997two}.
A major difference is that those closure problems for porous media only rely on periodic closure variables, whereas our result (\ref{eq: closure deviation fields quasi-developed region}) shows that actually exponentially decaying closure variables are inherently required to obtain exact closure for the quasi-developed heat transfer regime.

\newpage
\clearpage

\bibliography{elsarticle-template}

\end{document}